\documentclass[12pt]{article}
\usepackage[compress]{cite}
\usepackage{latexsym}
\usepackage{amsmath,amsfonts}
\usepackage{times}
\allowdisplaybreaks[4]

\catcode`\@=11

\newcount\hour
\newcount\minute
\newtoks\amorpm \hour=\time\divide\hour by 60\minute
=\time{\multiply\hour by 60 \global\advance\minute by-\hour}
\edef\standardtime{{\ifnum\hour<12 \global\amorpm={am}%
        \else\global\amorpm={pm}\advance\hour by-12 \fi
        \ifnum\hour=0 \hour=12 \fi
        \number\hour:\ifnum\minute<10
        0\fi\number\minute\the\amorpm}}
\edef\militarytime{\number\hour:\ifnum\minute<10 0\fi\number\minute}

\def\draftlabel#1{{\@bsphack\if@filesw {\let\thepage\relax
   \xdef\@gtempa{\write\@auxout{\string
      \newlabel{#1}{{\@currentlabel}{\thepage}}}}}\@gtempa
   \if@nobreak \ifvmode\nobreak\fi\fi\fi\@esphack}
        \gdef\@eqnlabel{#1}}
\def\@eqnlabel{}
\def\@vacuum{}
\def\marginnote#1{}
\def\draftmarginnote#1{\marginpar{\raggedright\scriptsize\tt#1}}
\def\draft{
        \pagestyle{plain}
        \overfullrule=2pt
        \oddsidemargin -.5truein
        \def\@oddhead{\sl \phantom{\today\quad\militarytime} \hfil
        \smash{\Large\sl DRAFT} \hfil \today\quad\militarytime}
        \let\@evenhead\@oddhead
        \let\label=\draftlabel
        \let\marginnote=\draftmarginnote
        \def\ps@empty{\let\@mkboth\@gobbletwo
        \def\@oddfoot{\hfil \smash{\Large\sl DRAFT} \hfil}
        \let\@evenfoot\@oddhead}
        \def\@eqnnum{(\theequation)\rlap{\kern\marginparsep\tt\@eqnlabel}%
        \global\let\@eqnlabel\@vacuum}  }

\newcommand{\rf}[1]{(\ref{#1})}
\renewcommand{\theequation}{\thesection.\arabic{equation}}
\renewcommand{\thefootnote}{\fnsymbol{footnote}}
\newcommand{\newsection}{   
\setcounter{equation}{0}\section}

\def\appendix#1{\addtocounter{section}{1}\setcounter{equation}{0}
\renewcommand{\thesection}{\Alph{section}}
\section*{Appendix \thesection\protect\indent \parbox[t]{11.15cm}{#1}}
\addcontentsline{toc}{section}{Appendix \thesection\ \ \ #1}}

\def\0psm{{\scriptscriptstyle (0)}}
\def\1psm{{\scriptscriptstyle (1)}}

\def\be{\begin{equation}}
\def\ee{\end{equation}}
\def\beq{\begin{eqnarray}}
\def\eeq{\end{eqnarray}}

\def\parline{\,\partial\kern -0.55em /\,\,}

\def\half{{\frac{1}{2}}}

\def\BB{{\cal B}}
\def\CC{{\cal C}}

\def\LL{{\cal L}}
\def\MM{{ M}}
\def\NN{{\cal N}}

\def\Cbf{{\bf C}}
\def\Dbf{{\bf D}}

\def\abf{{\bf a}}
\def\bbf{{\bf b}}
\def\cbf{{\bf c}}

\def\sbf{{\bf s}}

\def\ibf{{\bf i}}
\def\iibf{{\bf ii}}
\def\iiibf{{\bf iii}}

\def\Phibf{{\boldsymbol{\Phi}}}
\def\Pibf{{\boldsymbol{\Pi}}}

\def\alphabf{{\boldsymbol{\alpha}}}

\def\alphabfwh{{\widehat{\alphabf}}}

\def\isf{{\sf i}}
\def\jsf{{\sf j}}
\def\ksf{{\sf k}}
\def\lsf{{\sf l}}

\def\alphab{{\bar{\alpha}}}

\def\rhob{{\bar{\rho}}}
\def\etab{{\bar{\eta}}}
\def\zetab{{\bar{\zeta}}}

\def\FPsm{{\scriptscriptstyle FP}}
\def\Ksm{{\scriptscriptstyle{K}}}

\def\Ah{{\hat{A}}}
\def\Bh{{\hat{B}}}
\def\Ch{{\hat{C}}}

\def\Eh{{\hat{E}}}

\def\ah{{\hat{a}}}
\def\bh{{\hat{b}}}
\def\ch{{\hat{c}}}

\def\alphawh{{\widehat{\alpha}}}

\def\chib{{\bar{\chi}}}
\def\upsilonb{{\bar{\upsilon}}}

\def\Cb{\bar{C}}

\def\Lb{\bar{L}}

\def\fb{\bar{f}}

\def\chib{\bar{\chi}}

\def\(A)dS{{\rm (A)dS}}

\def\st{{\rm st}}
\def\md{{\rm md}}

\def\intrm{{\rm int}}
\def\gh{{\rm gh}}
\def\cvrm{{\rm cv}}
\def\Tsm{{\rm \scriptscriptstyle T}}

\def\mun{{\underline{m}}}

\def\noinbf#1{\noindent {\bf #1}}

\begin{document}


\begin{flushright}
FIAN-TD-2026-13 \hspace{1.8cm} {}~  \\
arXiv: 2607.02495 V4 [hep-th]  \\
\hspace{3.8cm}  {}~
\end{flushright}

\vspace{1cm}

\begin{center}

{\Large \bf BRST-BV approach to fields in Poincar\'e patch of AdS}

\vspace{2.5cm}

R.R. Metsaev%
\footnote{ E-mail: metsaev@lpi.ru
}

\vspace{1cm}

{\it Department of Theoretical Physics, Lebedev Physical Institute, \\
  Leninsky prospect 53, Moscow 119991, Russia }

\vspace{3.5cm}

{\bf Abstract}

\end{center}

We use the Poincar\'e parametrization of AdS space to develop a general BRST-BV approach for free fields. A general expression for the BRST-BV Lagrangian of fields with arbitrary masses and symmetry types is obtained. We apply this general framework to study totally symmetric massless, massive, and partially-massless fields with arbitrary integer spin and a continuous-spin field. For these fields, the constrained and unconstrained BRST-BV formulations are developed. We investigate both irreducible and reducible fields. In addition, we demonstrate the matching between the obtained BRST-BV Lagrangian and the metric-like Lagrangian formulated in terms of the modified de Donder derivative. Finally, a realization of AdS space symmetries is obtained within the space of fields and antifields entering the BRST-BV formulation. Modulo the choice of the Poincar\'e parametrization of AdS space, the Lagrangian and the entire approach are manifestly Lorentz invariant.

\newpage
\renewcommand{\thefootnote}{\arabic{footnote}}
\setcounter{footnote}{0}

\newsection{  Introduction }

The BRST approach \cite{Becchi:1974xu,Tyutin:1975qk} is one of the most powerful methods for studying modern quantum gauge field theory and string theory. In the context of gauge field theory, this framework is well-suited for implementing a manifestly Lorentz-covariant quantization procedure, as well as investigating ultraviolet divergences and renormalization. Furthermore, the BRST approach has proven invaluable for deriving the path-integral representation of the relativistically invariant S-matrix. Attempts to apply this methodology to the problem of constructing a manifestly Lorentz-covariant formulation of string field theory led to a novel and highly interesting development \cite{Siegel:1984wx,Siegel:1984xd,Hata:1985zu,Neveu:1985gx}. Specifically, extended versions of the BRST formalism incorporating antifields were obtained. Remarkably, these extended formulations turned out to be not only efficient for studying quantum field theory and string theory, but also powerful tools for constructing classical gauge field theories.
In this paper, the BRST approach with antifields \cite{Batalin:1981jr,Batalin:1984jr} will be referred to as the BRST-BV approach.

In Refs.\cite{Metsaev:2008ks,Metsaev:2009hp}, we studied totally symmetric massless and massive fields  with arbitrary integer spin in AdS space by using the metric-like approach. For these fields, we obtained two representations for the gauge-invariant Lagrangian: one in terms of a standard de Donder derivative and another in terms of a modified de Donder derivative. The latter representation, upon imposing the modified de Donder gauge, turned out to be convenient for analyzing the equations of motion. Namely, although the mass operator remains non-diagonal, the kinetic term becomes diagonal and is realized as that for a scalar field. This considerably simplifies the study of the equations of motion and, hence, the analysis of various problems related to the AdS/CFT correspondence. Note however, in this paper, we do not impose the modified de Donder gauge and deal with the gauge-invariant Lagrangian. One of the goals of this paper is to obtain a BRST-BV Lagrangian that is the counterpart of our gauge-invariant metric-like Lagrangian formulated in terms of the modified de Donder derivative.

In Refs.\cite{Metsaev:1999ui,Metsaev:2003cu}, we developed a general light-cone gauge formulation of relativistic dynamics for free fields in AdS space. The representation of the light-cone gauge Lagrangian obtained in Ref.\cite{Metsaev:2003cu} turns out to be universal, making it applicable to the study of both integer and continuous-spin fields with arbitrary symmetry types. The kernel of a mass-like term entering this Lagrangian is referred to as the AdS mass operator. This operator is expressed entirely in terms of spin operators that satisfy a set of algebraic relations, which we call defining equations. Finding the spin operators by solving these equations allows us to completely fix the light-cone gauge action. Because the defining equations are purely algebraic, constructing the light-cone gauge Lagrangian for free fields reduces to a purely algebraic problem. The main goal of this paper is to construct a BRST-BV counterpart of both the AdS mass operator and the defining equations. In doing so, we obtain a universal form for the BRST-BV Lagrangian of fields in the Poincar\'e patch of AdS space. What is very important for us is that this universal form is well suited for studying both irreducible fields and towers of fields. We then consider the specific cases of totally symmetric massless, massive, and partially-massless fields with arbitrary integer spin and a continuous-spin field. We demonstrate how our general results are realized for these simplest cases.

The paper is organized as follows.

In Sec.~\ref{sec-02}, we present our main results on the general form of the BRST-BV Lagrangian. In particular, we describe the general representation of the BRST charge and the AdS mass operator in terms of the spin operators. We also provide the defining equations for the spin operators that enter the BRST charge and the AdS mass operator. We derive the representation for the fourth-order Casimir operator of the $so(d,2)$ algebra in terms of these spin operators. Finally, we present a realization of the relativistic $so(d,2)$ symmetry algebra in a space of fields and anti-fields.

In Secs.~\ref{sec-03},\ref{sec-04} we apply our general results to the study of totally symmetric massless, massive, partially-massless fields with arbitrary integer spin, while, in Sec.~\ref{sec-05}, we consider a continuous-spin field. In particular, we describe the field content, present solution for the spin operators entering the AdS mass operator, and discuss both constrained and unconstrained formulations. In Sec.~\ref{sec-06}, we summarize our results and briefly discuss potential applications and generalizations.

In App.~A, we present our notation and conventions. In App.~B, we describe (anti)com\-mutators of the $osp(d,1)$ algebra, while in App.~C, we present an oscillator realization of this algebra. In App.~D, we first review representations for the metric-like gauge invariant Lagrangian  via standard and modified de Donder derivatives, and then  demonstrate how the Lagrangian via modified de Donder divergence matches the one obtained from the BRST-BV approach. In App.~E, we describe the spin operators appearing in the unconstrained formulations. In App.~F, we present the method of the $M\hbox{-}T$ pair for solving the basic equations for the spin operators.

\newsection{General BRST-BV Lagrangian} \label{sec-02}

\noinbf{Fields}. We start with the notation for the fields involved in our BRST-BV formulation in the Poincar\'e patch of AdS space. We prefer to use an index-free presentation. Therefore, we collect all fields (and anti-fields) into the generating field $\Phi$, which can be schematically presented as
\be \label{02062026-man-01}
\Phi = \Phi(x,z,\theta,\sbf)\,,
\ee
where $x=x^a$ and $z$ stand for the Poincar'e coordinates of AdS space, while $\theta$ is  a Grassmann coordinate $\theta$, $\theta^2=0$. The variable $\sbf$ is used to describe spin d.o.f fields in an index free way. The usual fields, which depend on the
space--time coordinates $x^a$, $z$ are obtainable by expanding the field
$\Phi$ \rf{02062026-man-01} in the $\theta$ and the
variables $\sbf$. Expansion in the $\theta$ is obvious
\beq
\label{02062026-man-02} \Phi = \phi(x,z,\sbf) + \theta \phi_*(x,z,\sbf)\,.
\eeq
For brevity, we refer to $\phi$ and $\phi_*$ as fields and antifields respectively. For a comment on the terminology in the literature, see Appendix A.

Depending on the choice of the variables $\sbf$ one can describe massless, massive, partially-massless AdS fields with integer spins and continuous-spin AdS fields. Our general results in this section are valid for all these fields with an arbitrary type of symmetry. An explicit description of the variables $\sbf$ will be given below when discussing totally symmetric fields.

\bigskip
\noinbf{BRST-BV Lagrangian}. The general
expression for the BRST-BV action we explore takes the form%
\footnote{If we ignore the dependence of $\Phi$ on the radial coordinate $z$, then the general form of the BRST-BV Lagrangian \rf{02062026-man-30},\rf{02062026-man-31} reduces to the one found for fields in flat space in the seminal works \cite{Siegel:1984wx,Siegel:1984xd,Hata:1985zu,Neveu:1985gx}.}
\beq
\label{02062026-man-30} && S = \int d^dx dz \, \LL\,, \qquad  \LL = \int d\theta\, \Phi^{\hat\dagger}\, Q\,  \Phi\,,\qquad  \Phi^{\hat\dagger} := \Phi^\dagger K\,,
\eeq
where the BRST operator $Q$ admits the general representation
\be \label{02062026-man-31}
Q = \theta (\Box - \MM^2 ) + M^{\eta a} \partial^a + M^\eta + \half M^{\eta\eta} \partial_\theta\,.
\ee
In \rf{02062026-man-31} and below, $\partial_\theta$ denotes the left derivative
of the Grassmann coordinate, $\partial_\theta = \partial/\partial\theta$, while
$\Box=\partial^a\partial^a$ is the d'Alembert operator in $R^{d-1,1}$.
The operators $M^2$, $M^{\eta a}$, $M^\eta$, and $M^{\eta\eta}$ are independent of the coordinates $x^a$, $\theta$ and their derivatives. These operators are realized as differential operators acting on the radial coordinate $z$ and the spin variables $\sbf$.  $\Phi^\dagger$ denotes the Hermitian conjugate of $\Phi$. The kernel $K$, $K^\dagger=K$, depends only on the spin variables and their derivatives.

As usually, BRST--BV action \rf{02062026-man-30} is invariant under the gauge
transformations
\beq \label{02062026-man-35}
&& \delta \Phi = Q \Xi\,,  \hspace{1cm} \Xi =  \Xi(x,z,\theta,\sbf)\,,
\eeq
where $\Xi$ denotes a generating field of gauge transformation
parameters. As usual, the BRST charge must be nilpotent, $Q^2=0$. A more detailed discussion of the nilpotent BRST charge and its hermiticity properties is given at the end of this section.

\noinbf{AdS mass operator and AdS spin operator}. From the above discussion,  we see that to find the BRST-BV action and the corresponding gauge
transformations we should fix the BRST operator. In turn, as seen from \rf{02062026-man-31}, to build the BRST operator, we should obtain a realization for the operators $M^2$, $M^{\eta a}$, $M^\eta$, and $M^{\eta\eta}$ in the space of $\Phi$. We now present our realization for all these operators.  The operators $M^{\eta a}$ and $M^{\eta\eta}$ turn out to be generators of the $osp(d,1|2)$ algebra and the realization for them is fixed by representation theory of this algebra.%
\footnote{As a side note, the origin of the auxiliary $osp$ algebra can be traced to the covariantization approach  proposed by Siegel and Zwiebach in their seminal paper \cite{Siegel:1986zi}.}
These operators act only on the spin variables $\sbf$. The difficulty lies in finding the operators $M^2$ and $M^\eta$. We proceed as follows.

First, we find the dependence of $M^2$ and $M^\eta$ on the radial coordinate $z$ and its derivative. Namely, it turns out that the operators $M^2$ and $M^\eta$ take the form
\beq
\label{02062026-man-41} && M^2 =  -\partial_z^2 + \frac{1}{z^2}A\,, \qquad  M^\eta =   M^{\eta z}\partial_z + \frac{1}{2z}  [M^{\eta z},A]\,,
\eeq
$\partial_z:=\partial/\partial z$, where we introduce new operators $A$ and $M^{\eta z}$ which act only on the spin variables $\sbf$. In other words, relations \rf{02062026-man-41} exhaust all dependence of the operators $M^2$, $M^\eta$ on the radial coordinate $z$ and its derivative.  The operator $M^{\eta z}$ turns out to be the generator of the $osp(d,1|2)$ algebra. The operator $A$ is referred to as the AdS mass operator.

Second, we obtain the following general representation for the AdS mass operator:
\beq
\label{02062026-man-45}  A = \CC_2 + \frac{d^2-1}{4} + \half M^{\ah\bh}M^{\bh\ah}  - M^{\Ah\Bh}M^{\Bh\Ah} + 2 B^z \,,
\eeq
where $\CC_2$ is the 2nd-order Casimir of the $so(d,2)$ algebra which describes the  relativistic symmetries of $AdS_{d+1}$. The operators $M^{\Ah\Bh}$, $M^{\ah\bh}$  are generators of the respective $osp(d,1|2)$ and $osp(d-1,1|2)$ algebras, while the operator $B^z$ is the $z$-component of the vector operator $B^\Ah$ of the $osp(d,1|2)$ algebra.
The indices $\Ah$, $\Bh$ and $\ah$, $\bh$ are vector indices of the respective  $osp(d,1|2)$ algebra and $osp(d-1,1|2)$ sub-algebra, while $B^\Ah$ is decomposed as $B^\Ah= (B^\ah,B^z)$.  A detailed description of the relevant algebras and the notation are relegated to Apps.~A, B.

Third, we find that the operator $B^\Ah$ should satisfy the defining equations of our approach, which constitute the central result of the present paper,
\beq
\label{02062026-man-50}  && \hspace{-2cm} [B^\Ah,B^\Bh\} = \Bigl( \CC_2 - \half M^{\Ch\Eh}M^{\Eh\Ch} + \frac{d^2 - 3d + 4}{2}  \Bigr) M^{\Ah \Bh} - (M^3)^{[\Ah|\Bh\}}\,,
\\
\label{02062026-man-51} && (M^3)^{[\Ah|\Bh\}} := \half (M^3)^{\Ah\Bh} -   \half  (-)^{\varepsilon_\Ah\varepsilon_\Bh} (M^3)^{\Bh\Ah}\,,
\nonumber\\
&& (M^3)^{\Ah\Bh} := M^{\Ah\Ch} M^{\Ch \Eh} M^{\Eh \Bh}\,,
\eeq
where operators $M^{\Ah\Bh}$ are generators of the $osp(d,1|2)$ algebra, while $[X,Y\}$ is a graded commutator (for details of the notation, see Apps.~A, B). (Anti)commutators between $B^\Ah$ and the generators of the $osp(d,1|2)$ algebra are given by
\beq
&& [B^\Ah,M^{\Bh\Ch}\} = \eta^{\Ah\Bh} B^\Ch - (-)^{\varepsilon_\Bh\varepsilon_\Ch} \eta^{\Ah\Ch} B^\Bh\,.
\eeq
(Anti)commutators of the $osp(d,1|2)$ algebra are described in App.~B.
The operators $B^\Ah$ and $M^{\Ah\Bh}$ will be referred to as spin operators.

\medskip
The following remarks are in order.

\noindent

\noinbf{i}) Relations given in \rf{02062026-man-41}-\rf{02062026-man-50} provide a complete description of the BRST operator $Q$. It follows from \rf{02062026-man-41} and \rf{02062026-man-45} that knowing the spin operators $M^{\Ah\Bh}$ and $B^\Ah$ is sufficient to uniquely determine the BRST operator. The expressions for the spin operators $M^{\Ah\Bh}$ are determined by the representation theory of the $osp(d,1|2)$ algebra, whereas the operators $B^\Ah$ are fixed by equations \rf{02062026-man-50}. In other words, the problem of finding the explicit form of the BRST operator $Q$ reduces to solving the algebraic equations \rf{02062026-man-50}.

\noinbf{ii}) As follows from \rf{02062026-man-45}, the AdS mass operator $A$ is expressed entirely in terms of the spin operators $B^\Ah$, $M^{\Ah\Bh}$, and the second-order Casimir operator $\CC_2$ of the $so(d,2)$ algebra. In other words, the operators $B^\Ah$ and $M^{\Ah\Bh}$ are the main ingredients of our approach.

\noinbf{iii}) The operator $\CC_2$ is the counterpart of the mass-squared operator for fields in flat space. In flat space, the mass-squared operator commutes with all the spin operators. For fields in AdS, the operator $\CC_2$ also commutes with the spin operators $B^\Ah$ and $M^{\Ah\Bh}$.

\noinbf{iv)} The alternative form of the operator $A$ \rf{02062026-man-45} and the operator $M^\eta$ \rf{02062026-man-41} is given by
\beq
&& A = \CC_2 + \frac{d^2-1}{4}  - \half M^{\ah\bh} M^{\bh\ah} + 2 M^{z\ah} M^{z\ah} + 2B^z\,,
\nonumber\\
&& M^\eta =   M^{\eta z}\partial_z + \frac{1}{z} B^\eta - \frac{1}{2z}\big( M^{\eta \ah}M^{\ah z} + M^{z\ah} M^{\ah\eta}\big)\,.
\eeq

\noinbf{v)} If equations \rf{02062026-man-50} can be solved in the strong (weak) sense, then the relation $Q^2=0$ holds also in the strong (weak) sense.

\noinbf{vi)} The general strategy for deriving relations and equations \rf{02062026-man-41}-\rf{02062026-man-50} is as follows. We postulate: \abf) the general expression for the BRST charge \rf{02062026-man-31}, where the operators $M^{\eta a}$ and $M^{\eta\eta}$ are generators of the $osp(d,1|2)$ algebra; \bbf) the expression for the operator $M^2$ given in \rf{02062026-man-41}, where $A$ is an operator independent of the coordinates $x^a$, $z$, $\theta$ and their derivatives; and \cbf) the expression for the operator $M^\eta = M^{\eta z} \partial_z + \frac{1}{z} L^\eta$, where the operator $M^{\eta z}$ is a generator of the $osp(d,1|2)$ algebra, while $L^\eta$ is an operator independent of the coordinates $x^a$, $z$, $\theta$ and their derivatives.
The expressions for $L^\eta$, $A$, as well as equations \rf{02062026-man-41}-\rf{02062026-man-50} are then obtained by requiring $Q^2=0$ and the invariance of action \rf{02062026-man-30} under the relativistic $so(d,2)$ symmetry.

\medskip
\noinbf{Realization of AdS space  symmetries}. Relativistic symmetries of fields propagating in $AdS_{d+1}$ are described by the $so(d,2)$ algebra. As we use the Poincar\'e parametrization of $AdS_{d+1}$, only symmetries of the Lorentz sub-algebra $so(d-1,1)$ are manifest in the framework of our approach. Therefore, in order to complete our BRST-BV formulation, we should provide a realization of symmetries of the $so(d,2)$ algebra in the space of $\Phi$ \rf{02062026-man-01}. As usual, when using the Poincar\'e parametrization of $AdS_{d+1}$, it is convenient to represent the generators of the $so(d,2)$ algebra in terms of the Poincar\'e translation generators $P^a$, dilatation generator $D$, conformal boost generators $K^a$, and generators of the Lorentz sub-algebra $so(d-1,1)$ denoted by $J^{ab}$. We then use the following form for commutators of the $so(d,2)$ algebra generators
{\small
\beq
\label{02062026-man-75} && {}[D,P^a]=-P^a\,,
\hspace{2.5cm}
[P^a,J^{bc}] =\eta^{ab}P^c -\eta^{ac}P^b
\,,
\nonumber\\
&& [D,K^a]=K^a\,,
\hspace{2.7cm}
[K^a,J^{bc}] = \eta^{ab}K^c - \eta^{ac}K^b\,,
\\
&& [P^a,K^b] = \eta^{ab}D - J^{ab}\,,
\hspace{1.2cm}
[J^{ab},J^{ce}] = \eta^{bc} J^{ae} + 3\hbox{ terms}.
\nonumber
\eeq
}

For fields in $AdS_{d+1}$, a general realization of commutators \rf{02062026-man-75} is given by
\beq
\label{02062026-man-77} && P^a = \partial^a\,, \hspace{2cm}  J^{ab} = x^a\partial^b - x^b \partial^a + M^{ab}\,,
\nonumber\\
&& D = x^a\partial^a + \Delta\,,\hspace{1cm}  K^a =  - \half x^b x^b \partial^a + x^a D + M^{ab} x^b + R^a\,, \qquad
\eeq
where $M^{ab}$ is a spin operator of the Lorentz sub-algebra $so(d-1,1)$. The operators $\Delta$ and $R^a$ \rf{02062026-man-77} being explicitly independent of the coordinates $x^a$ are realized as differential operators acting on the space-time coordinates $x^a$, $z$ and oscillators $\sbf$ entering field $\Phi$ \rf{02062026-man-01}. Notably, constructing the operator $R^a$ constitutes the main technical challenge in realizing the relativistic symmetries within our framework. The resulting expressions for $\Delta$ and $R^a$ take the form
\beq
&& \Delta = z\partial_z + 2\theta\partial_\theta - M^{\eta\rho} +  \frac{d-1}{2}\,, \qquad  R^a = - \half z^2 \partial^a - z M^{za} + 2\theta M^{\rho a}\,,
\eeq
where $M^{za}$, $M^{\eta\rho}$, $M^{\rho a}$ are generators of the $osp(d,1|2)$ algebra. Note that using Faddeev-Popov ghost number operator \rf{20062026-man-50}, we can represent $\Delta$ as $\Delta = z\partial_z + \theta\partial_\theta + N_\FPsm  + \frac{d-1}{2}$.

\noinbf{4th order Casimir operator}. Let us introduce an operator constructed out of the spin operators and denote it as $\CC_4$,
{\small
\be \label{02062026-man-87}
\CC_4 = B^\Ah B^\Ah + \half \big(\CC_2 + \frac{d^2-3d+4}{4}\big)M^{\Ah\Bh}M^{\Bh\Ah} - \frac{1}{8} (M^{\Ah\Bh}M^{\Bh\Ah})^2 - \frac{1}{4} M^{\Ah\Bh}M^{\Bh\Ch}M^{\Ch\Eh}M^{\Eh\Ah}\,.\qquad
\ee
}
\!Regarding the operator $\CC_4$ in \rf{02062026-man-87} we make the following observations. First,  $\CC_4$ commutes with all spin operators $B^\Ah$, $M^{\Ah\Bh}$.  Second, $\CC_4$ \rf{02062026-man-87} looks like the $osp(d,1|2)$ covariantization of the light-cone gauge expression for the 4th-order Casimir operator of the $so(d,2)$ algebra given in formula (2.17) of Ref.\cite{Metsaev:2019opn}. Third, for all totally symmetric fields under investigation, we verified that the eigenvalues of $\CC_4$ \rf{02062026-man-87} coincide with the eigenvalues of the 4th-order Casimir operator $\CC_4$ of the $so(d,2)$ algebra prescribed by group theory. All in all, we identify $\CC_4$ in \rf{02062026-man-87} as the BRST-BV form of the 4th-order Casimir operator $\CC_4$ of the $so(d,2)$ algebra.

\noinbf{Totally symmetric fields}. For the case of totally symmetric fields some relations given above are simplified. Namely, for such fields, the spin operator $M^{\Ah\Bh}$ satisfies the following identity:
\beq
\label{02062026-man-90}  && (M^3)^{[\Ah|\Bh\}} = \half \big(M^{\Ch\Eh}M^{\Eh\Ch} + (d-3)(d-4)\big)M^{\Ah\Bh}\,.
\eeq
The explicit expression for $M^{\Ah\Bh}$ can be chosen as in \rf{21062026-man-50}. In view of \rf{02062026-man-90}, the equation for the spin operator $B^\Ah$ \rf{02062026-man-50} and the expression for the operator $\CC_4$ \rf{02062026-man-87} take the simpler form given by
\beq
\label{02062026-man-92}  && [B^\Ah,B^\Bh\} = \big( \CC_2 - M^{\Ch\Eh}M^{\Eh\Ch} + 2d - 4\big) M^{\Ah \Bh}\,,
\\
\label{02062026-man-93} && \CC_4 = B^\Ah B^\Ah + \half \big(\CC_2 +  d -2\big)M^{\Ah\Bh}M^{\Bh\Ah} - \frac{1}{4} (M^{\Ah\Bh}M^{\Bh\Ah})^2\,.
\eeq

For the reader's convenience,  we summarize here how the relation $Q^2=0$ and the hermiticity properties of $Q$ are realized for totally symmetric fields.

{\it Massless field with integer spin}. In the constrained and unconstrained formulations, the relation $Q^2 = 0$ holds in the strong sense. Within these frameworks, we choose either $K=1$ or $K\ne 1$. For both formulations, the hermiticity condition $(K Q)^\dagger =  K Q$ holds strongly.

{\it Massive, partially-massless fields with integer spin, and continuous-spin fields}. In the constrained formulations, the relation $Q^2 = 0$ holds only on the subspace of  $osp(d,1|2)$-traceless fields, whereas in the unconstrained formulations, $Q^2=0$ holds strongly.
In the constrained formulations, we use $K=1$ and the hermiticity condition $\Phi^\dagger Q^\dagger \Phi =  \Phi^\dagger Q \Phi$, whereas in the unconstrained formulations, $K\neq1$ in general and the hermiticity condition $(KQ)^\dagger = KQ$ holds strongly.

We now consider various totally symmetric fields in turn.

\newsection{ Massless fields } \label{sec-03}

Equations for higher-spin massless fields proposed in  Refs.\cite{Vasiliev:1990en,Vasiliev:2003ev} triggered interest in the active study of higher-spin fields both in AdS and flat spaces.
Recent studies on this topic can be found, e.g., in Refs.\cite{Tatarenko:2024csa,Tatarenko:2025krq, Didenko:2025xca,Faliakhov:2026yum,Didenko:2026nip, Korybut:2025vdn,Gelfond:2025alv, Didenko:2021vdb,Skvortsov:2025ohi,Serrani:2026azw, Serrani:2025owx,Serrani:2026dbs,Guarini:2026vds,Cho:2026mjp, Misuna:2026has,Boulanger:2026wle,Buchbinder:2025yef} .
Various BRST studies of massless AdS fields are available, e.g., in Refs.\cite{Bengtsson:1990un,Buchbinder:2001bs,Alkalaev:2009vm}. Interesting ambient-space BRST variational calculus was developed in Ref.\cite{Bekaert:2023uve}.
Various approaches to interacting light-cone gauge AdS fields were considered in Refs.\cite{Metsaev:2018xip,Kozaki:2025jrj,Kozaki:2026rdc}.  The use of light-cone gauge approach for the study of AdS/CFT correspondence may be found, e.g., in Refs.\cite{Skvortsov:2018uru,deMelloKoch:2024juz,deMelloKoch:2024nfj, deMelloKoch:2024otg,deMelloKoch:2024ewt,deMelloKoch:2024aqh}.

\subsection{\large Constrained formulation}

\noinbf{Field content}. The kernel $K$ is fixed to be $K=1$, while the spin variables in \rf{02062026-man-01} are fixed to be
\beq
\label{03062026-man-01} && \sbf = \alpha^\Ah\,, \qquad \alpha^\Ah = \alpha^A\,, \eta\,, \rho\,.
\eeq
The $\alpha^A$ are Grassmann-even creation operators, while  $\eta$, $\rho$ are Grassmann-odd creation operators. Expansion of $\phi$ and $\phi_*$ in the Grassmann-odd oscillators takes the form
\beq
\label{03062026-man-04}  && \phi = \phi_0 + \eta \phi_\eta + \rho \phi_\rho + \rho\eta \phi_{\eta\rho}\,, \qquad  \phi_* = \phi_{*0} + \eta \phi_{*\eta} + \rho \phi_{*\rho} + \rho\eta \phi_{*\eta\rho}\,,
\eeq
where the fields $\phi_0$, $\phi_\eta$, $\phi_\rho$, $\phi_{\eta\rho}$ depend only on the oscillators $\alpha^A$ and take the following form
{\small
\beq
\label{03062026-man-10}  &&  \phi_0 = \frac{1}{s!} \alpha^{A_1} \ldots \alpha^{A_s} \phi_0^{A_1\ldots A_s}\,,
\nonumber\\
&&   \phi_{\eta,\rho} = \frac{1}{(s-1)!}  \alpha^{A_1} \ldots \alpha^{A_{s-1}} \phi_{\eta,\rho}^{A_1\ldots A_{s-1}}\,,
\nonumber\\
&& \phi_{\eta\rho} =  \frac{1}{(s-2)!} \alpha^{A_1} \ldots \alpha^{A_{s-2}} \phi_{\eta\rho}^{A_1\ldots A_{s-2}}\,.
\eeq
}
\!The component fields $\phi^{A_1\ldots A_n}$ appearing in \rf{03062026-man-10}, are scalar, vector, and totally symmetric tensor fields under the $so(d,1)$ Lorentz algebra. Needless to say that $\phi_*$ takes the same expansion in the oscillators as $\phi$. It is easy to see that $\Phi$ satisfies the following homogeneity condition:
\beq
\label{03062026-man-15} \big( N_\alphabfwh - s) \Phi = 0\,,
\eeq
which tells us that $\Phi$ is degree-$s$ homogeneous polynomial in the oscillators $\alpha^\Ah$.

To obtain a description for an irreducible massless field we impose the tracelessness constraint,
\beq
\label{03062026-man-20} \alphab^\Ah \alphab^\Ah \Phi = 0\,.
\eeq
The tracelessness constraint \rf{03062026-man-20} is invariant under the $osp(d,1|2)$ algebra. In view of the constraint \rf{03062026-man-20}, the formulation we discuss is referred to as the constrained formulation. Note that $\Xi$ appearing in the gauge transformation \rf{02062026-man-35} should also satisfy the tracelessness constraint.

\noinbf{Solution for spin operators}. The spin operators $M^{\Ah\Bh}$ for totally symmetric fields are well known and are given in \rf{21062026-man-50}, App.~C, while the spin operator $B^\Ah$ vanishes. This leads to a simpler AdS mass operator $A$ compared to the one in \rf{02062026-man-45} and an unexpected representation for the 2nd-order Casimir operator $\CC_2$ of the $so(d,2)$ algebra. All in all, our result is summarized by the relations
\beq
\label{03062026-man-25} && B^\Ah = 0 \,,
\\
\label{03062026-man-26} && A =  \half M^{\ah\bh} M^{\bh\ah} + \frac{(d-3)(d-5)}{4}\,,
\\
\label{03062026-man-27} && \CC_2 =  M^{\Ah\Bh} M^{\Bh\Ah} -2d +4\,.
\eeq
We now make comments on the derivation and various properties of our solution \rf{03062026-man-25}-\rf{03062026-man-27}.

\noinbf{i)} Our solution leads to the BRST charge satisfying $Q^2=0$, $Q^\dagger = Q$ in the strong sense.

\noinbf{ii)} In deriving our solution, we do not use constraint \rf{03062026-man-20} and our solution satisfies the basic equations \rf{02062026-man-92} in the strong sense. For the spin operators, hermiticity conditions \rf{19062026-man-37} with $K=1$ are also realized in the strong sense. Constraint \rf{03062026-man-20} is required to describe the irreducible spin-$s$ massless field. If we ignore constraint \rf{03062026-man-20}, then our solution describes a spin-$s$ triplet massless field. We recall that the massless spin-$s$ triplet field  is decomposed into a chain of irreducible massless fields with spins $s,s-2,\ldots,(0,1)$.
Various interesting Lorentz covariant studies of the triplet field may be found, e.g., in Refs.\cite{Bengtsson:1986ys,Pashnev:1989gm,Francia:2002pt,Sagnotti:2003qa, Sorokin:2008tf,Sorokin:2018djm,Campoleoni:2012th, Francia:2016weg}.

\noinbf{iii)} Relation \rf{03062026-man-25} can easily be understood by noting that $B^\Ah$ should respect constraint \rf{03062026-man-15}, $[B^\Ah,N_\alphabfwh]=0$. Obviously, by using only one oscillator $\alpha^\Ah$, it is impossible to build $B^\Ah$ that commutes  with $N_\alphabfwh$.
From Ref.\cite{Metsaev:2003cu}, we learn that, for totally symmetric massless integer spin field, the light-cone gauge counterpart of the operator $B^\Ah$ is equal to zero and hence  relation \rf{03062026-man-25} is in agreement with the light-cone gauge analysis.

\noinbf{iv)} Substituting \rf{03062026-man-25} into \rf{02062026-man-92} leads to a somewhat unexpected solution for $\CC_2$ \rf{03062026-man-27} which tells us that, for the case under consideration, $\CC_2$ is expressed in terms of the generators of the $osp(d,1|2)$ algebra. We now verify that relation \rf{03062026-man-27} leads to the correct eigenvalue of $\CC_2$. To this end, we recall that on the one hand, the eigenvalue of $\CC_2$ for a massless totally symmetric field in $AdS_{d+1}$ is given by
\beq
\label{03062026-man-35} && \langle\CC_2\rangle = 2(s-1)(s+d-2).
\eeq
On the other hand, using \rf{03062026-man-15}, \rf{03062026-man-20}, we find
{\small
\be \label{03062026-man-36}
M^{\Ah\Bh} M^{\Bh\Ah}\Phi = 2s(s+d-3)\Phi\,,\qquad  M^{\Ah\Bh} M^{\Bh\Ah} = 2 N_\alphabfwh(N_\alphabfwh + d-3) -  2\alpha^\Ah\alpha^\Ah \alphab^\Bh \alphab^\Bh\,,\qquad
\ee
}
\!where we show the expression for $M^{\Ah\Bh}M^{\Bh\Ah}$ to demonstrate that in order to diagonalize $M^{\Ah\Bh} M^{\Bh\Ah}$ we should use tracelessness constraint \rf{03062026-man-20}. Using \rf{03062026-man-36} in \rf{03062026-man-27}, we obtain \rf{03062026-man-35}.

\noinbf{v)} Plugging $B^\Ah$, $M^{\Ah\Bh}$ into \rf{02062026-man-93} and using \rf{03062026-man-36}, we verify, as promised, that the eigenvalue of $\CC_4$ \rf{02062026-man-93} coincides with the eigenvalue of the $so(d,2)$ algebra Casimir operator $\CC_4$  corresponding to a massless spin-$s$ field. In App.~D, we recall the eigenvalues of $\CC_4$ for all fields under investigation.

\noinbf{vi)} It is instructive to relate our BRST-BV Lagrangian to the standard metric-like Lagrangian. To this end, we note that all fields having non-zero ghost numbers can be eliminated from consideration by using a suitable gauge condition. After this, our BRST-BV Lagrangian leads to a metric-like, gauge-invariant Lagrangian which exactly coincides with the one formulated in terms of the modified de Donder derivative in Ref.\cite{Metsaev:2008ks}.  For details of the matching, see App.~D.

\noinbf{vii}) Solution \rf{03062026-man-25}-\rf{03062026-man-27} has a remarkable property: it does not depend on the spin $s$ and is valid for arbitrary $s$. This implies that our solution can be used not only for an irreducible massless field but also for a finite or infinite tower of massless fields. We refer to solutions with such a property as uniform solutions. To clarify what has just been said, we use the notation $\Phi_s$ for a field satisfying \rf{03062026-man-15}, \rf{03062026-man-20} and introduce the following tower of massless fields:
\beq
\label{03062026-man-38} && \Phibf = \sum_s \Phi_s\,.
\eeq
We now note that solution \rf{03062026-man-36} remains valid for the tower of fields \rf{03062026-man-38}.

\noinbf{viii)} To some extent, formula \rf{03062026-man-27} resembles the relation for the 2nd-order Casimir operator of the Poincar\'e algebra in string theory. In string theory, the  Casimir operator is equal to the square of mass operator which is quadratic in the string  oscillators. In our case, we have only one oscillator $\alpha^\Ah$, while the $\CC_2$ is quartic in this oscillator.

\noinbf{ix)} Our fields are assumed to be singlets of an internal symmetry algebra. Incorporation of internal symmetries can be done, e.g., as in Refs.\cite{Konstein:1989ij,Metsaev:1991nb,Skvortsov:2020wtf,Skvortsov:2018jea}.

\subsection{\large Unconstrained formulation}

\noinbf{Field content}. The spin variables entering generating function $\Phi$ \rf{02062026-man-01} are fixed to be
\beq
\label{03062026-man-01a} && \sbf = \alpha^\ah\,, \chi\,,  \qquad \alpha^\ah = \alpha^a\,, \eta\,, \rho\,,\quad \chi^2 = 0 \,,
\eeq
where $\alpha^a$, $\chi$ are Grassmann-even creation operators, while $\eta$, $\rho$ are Grassmann-odd creation operators. The oscillator $\alpha^\ah$ is a vector under the $osp(d-1,1|2)$ subalgebra. An expansion of $\phi$ and $\phi_*$ in the Grassmann-odd oscillators $\eta$, $\rho$ takes the form given in \rf{03062026-man-04}, where now the fields $\phi_0$, $\phi_\eta$, $\phi_\rho$, $\phi_{\eta\rho}$ depend on the oscillators $\alpha^a$, $\chi$ and take the following explicit form:
{\small
\beq
&& \hspace{-0.8cm}\phi_0 = \phi_{0\,0} + \chi \phi_{0\,1} \,, \quad \phi_\eta = \phi_{\eta\,0} + \chi \phi_{\eta\,1} \,, \quad \phi_\rho = \phi_{\rho\,0} + \chi \phi_{\rho\,1} \,, \quad \phi_{\eta\rho} = \phi_{\eta\rho\,0} + \chi \phi_{\eta\rho\,1} \,, \qquad
\\
\label{03062026-man-10a}  &&  \phi_{0\, k} := \frac{1}{(s-k)!} \alpha^{a_1} \ldots \alpha^{a_{s-k}} \phi_{0\, k}^{a_1\ldots a_{s-k}}\,,
\nonumber\\
\label{03062026-man-11a}&&   \phi_{\eta,\rho\,k} := \frac{1}{(s-1-k)!}  \alpha^{a_1} \ldots \alpha^{a_{s-1-k}} \phi_{\eta,\rho\, k}^{a_1\ldots a_{s-1-k}}\,,
\nonumber\\
&& \phi_{\eta\rho\,k} :=  \frac{1}{(s-2-k)!} \alpha^{a_1} \ldots \alpha^{a_{s-2-k}} \phi_{\eta\rho\,k}^{a_1\ldots a_{s-2-k}}\,,\qquad k=0,1,
\eeq
}
\!where, in \rf{03062026-man-10a}, the shortcut $\phi_{\eta,\rho\,k}$ stands for the fields  $\phi_{\eta\,k}$ and $\phi_{\rho\,k}$ while the component fields $\phi^{a_1\ldots a_n}$ are scalar, vector, and totally symmetric tensor fields under the $so(d-1,1)$ subalgebra. All tensor fields are traceful and there are no other constraints on the fields, meaning that we deal with the unconstrained formulation. It is easy to see that the field $\Phi$ satisfies the following homogeneity condition
\beq
\label{03062026-man-15ax} \big( N_\alphawh + N_\chi- s) \Phi = 0\,,
\eeq
which tells us that $\Phi$ is degree-$s$ homogeneous polynomial in the oscillators $\alpha^\ah$, $\chi$.

\noinbf{Solution for spin operators}. All relations  \rf{03062026-man-25}-\rf{03062026-man-27} hold true for the unconstrained formulation. Spin operators \rf{03062026-man-25} also largely remain valid for this formulation. The only required modification is related to a new representation for the spin operator $M^{z\ah}$. The generators $M^{\Ah\Bh}$ of the $osp(d,1|2)$ algebra are decomposed into $M^{\ah\bh}$, $M^{z\ah}$, where $M^{\ah\bh}$ are the generators of the $osp(d-1,1|2)$ subalgebra, while $M^{z\ah}$ are boost generators in the radial $z$-direction. We now discuss two equivalent representations corresponding to the choices $K\ne 1$ and $K=1$.

\noindent {\it Case $K\ne 1$}. For this case, the operator $K$ and spin operators  $M^{\ah\bh}$ and $M^{z\ah}$ take the forms given in relations \rf{23062026-man-01}, \rf{23062026-man-05} of App.~E.
An advantage of the proposed formulation is that, as seen from relations \rf{23062026-man-05} of App.~E, all the spin operators and hence the BRST-BV operator are realized as polynomials of the oscillators. The relations $Q^2=0$, $(KQ)^\dagger= KQ$, and the hermiticity conditions for the spin operators  given in \rf{19062026-man-37} are realized strongly.

\noindent {\it Case $K =1$}. For this case, the spin operators  $M^{\ah\bh}$ and $M^{z\ah}$ take the forms given in relations \rf{23062026-man-30} of App.~E. Here, the spin operators are realized in terms of dressed oscillators. An advantage of the proposed formulation is that the relation $Q^2=0$ is realized strongly and the hermiticity property takes the desired simple strong form  $Q^\dagger=Q$.  For $B^\Ah$ and $M^{\Ah\Bh}$, hermiticity conditions \rf{19062026-man-37} with $K=1$ also take the desired simple strong form.

The following remarks are in order.

\noinbf{i)} We note the following strong relation for the square of the spin operators $M^{\Ah\Bh}$:
\be
\label{03062026-man-36a} M^{\Ah\Bh} M^{\Bh\Ah} = 2(N_\alphawh+N_\chi)(N_\alphawh + N_\chi + d-3)\,.
\ee
Relation \rf{03062026-man-36a} holds true for both cases discussed above.
Comparing \rf{03062026-man-36} and \rf{03062026-man-36a}, we note the absence of the $\alpha^\Ah\alpha^\Ah \alphab^\Bh \alphab^\Bh$-term in \rf{03062026-man-36a}. This implies that, in the unconstrained formulation, to diagonalize $M^{\Ah\Bh} M^{\Bh\Ah}$ we do not need to use the tracelessness constraint.

\noinbf{ii)} Throughout this paper, we derive unconstrained formulations directly from the constrained ones. Specifically, solving tracelessness constraint \rf{03062026-man-20} allows us to eliminate the oscillator $\alpha^z$. We perform this elimination in the same manner as the light-cone gauge approach in App.~B of Ref.\cite{Metsaev:1999ui}  (see relations  B7--B12 therein).

\section{ \large Massive and partially-massless fields }\label{sec-04}

In view of potentially interesting applications in AdS string theory, massive AdS fields remain the focus of current research. For convenience, we present an inevitably   incomplete list of references on this topic. Lagrangian metric-like and frame-like formulations of massive fields were developed in Refs.\cite{Zinoviev:2001dt,Zinoviev:2008ze,Ponomarev:2010st}. The BRST studies of massive fields may be found in Refs.\cite{Buchbinder:2005ua,Buchbinder:2006ge}. Various interesting recent studies of massive field may be found in Refs.\cite{Fegebank:2024yft,Buchbinder:2026oou}.

Partially-massless fields \cite{Deser:2001xr} also have been actively investigated during the last decade. Various metric-like and frame-like Lagrangian formulations of a partially-massless field were discussed in Refs.\cite{Zinoviev:2001dt,Skvortsov:2006at}. Partially-massless fields in the light-cone gauge and the BRST formalism in $AdS_4$, formulated in terms of bosonic spinor-like oscillators, were investigated in Refs.\cite{Metsaev:2022ndg,Buchbinder:2026cmo}. The study of the chiral partially-massless field may be found in Ref.\cite{Basile:2022mif}. Supersymmetric partially-massless fields were considered in Refs.\cite{Garcia-Saenz:2018wnw,Bittermann:2020xkl,Buchbinder:2019olk}. Steps towards higher-spin gravity involving partially-massless fields are presented in Refs.\cite{Joung:2015jza,Brust:2016zns,Brust:2016xif}. The (non-)unitarity of partially-massless fields was studied in Ref.\cite{Letsios:2022tsq,Letsios:2023qzq}. Interacting partially-massless fields are considered, for instance, in Refs.\cite{Joung:2012rv,Joung:2019wwf,Boulanger:2012dx,Zinoviev:2025tpx}.

We now discuss totally symmetric massive, partially-massless fields in $AdS_{d+1}$, $d\geq 3$. We develop constrained and unconstrained formulations for such fields. We discuss the formulations in turn.

\subsection{\large Constrained formulation}

\noinbf{Field content}. The kernel $K$ is fixed to be $K=1$, while the spin variables in \rf{02062026-man-01} are fixed to be
\beq
\label{07062026-man-01} && \sbf := \alpha^\Ah\,, \zeta\,, \qquad \qquad \alpha^\Ah: = \alpha^A\,, \eta\,, \rho\,, \qquad
\eeq
where $\alpha^A$, $\zeta$ are Grassmann-even creation operators, while $\eta$, $\rho$ are Grassmann-odd creation operators.  Expansion of $\phi$ and $\phi_*$ in $\eta$, $\rho$ takes the form
\beq
\label{07062026-man-03} && \phi = \phi_0 + \eta \phi_\eta + \rho \phi_\rho + \rho\eta \phi_{\eta\rho}\,, \qquad  \phi_* = \phi_{*0} + \eta \phi_{*\eta} + \rho \phi_{*\rho} + \rho\eta \phi_{*\eta\rho}\,,
\eeq
where the fields $\phi_0$, $\phi_\eta$, $\phi_\rho$, $\phi_{\eta\rho}$ depend only on the oscillators $\alpha^A$, $\zeta$  and take the following form
{\small
\beq
\label{07062026-man-05} &&  \phi_0 = \sum_{n=s-t}^s \frac{\zeta^{s-n}\alpha^{A_1} \ldots \alpha^{A_n}}{n!\sqrt{(s-n)!}} \phi_0^{A_1\ldots A_n}\,,
\nonumber\\
&&   \phi_{\eta,\rho} = \sum_{n=s-1-t\atop n\geq 0}^{s-1} \frac{\zeta^{s-1-n}\alpha^{A_1} \ldots \alpha^{A_n}}{n!\sqrt{(s-1-n)!}} \phi_{\eta,\rho}^{A_1\ldots A_n}\,,
\nonumber\\
&& \phi_{\eta\rho} = \sum_{n=s-2-t\atop n \geq 0 }^{s-2} \frac{\zeta^{s-2-n}\alpha^{A_1} \ldots \alpha^{A_n}}{n!\sqrt{(s-2-n)!}}\phi_{\eta\rho}^{A_1\ldots A_n}\,,
\nonumber\\
&& \hspace{1cm}t = s \hbox{ for massive field}; \qquad t = 1,\ldots,s-1\,, \hbox{for partially-massless fields}.\qquad
\eeq
}
\!The fields $\phi^{A_1\ldots A_n}$ appearing in \rf{07062026-man-05}, are scalar, vector, and totally symmetric tensor fields under the $so(d,1)$ Lorentz algebra.
The field $\Phi$ satisfies the homogeneity condition
\beq
\label{07062026-man-07} \big( N_\alphabfwh + N_\zeta - s) \Phi = 0\,,
\eeq
which tells us that $\Phi$ is degree-$s$ homogeneous polynomial in the variables $\alpha^\Ah$ and $\zeta$. In \rf{07062026-man-07} we use the notation given in \rf{20062026-man-10}, App.~A.

To obtain a description for an irreducible massive/partially-massless field, we impose the tracelessness constraint
\beq
\label{07062026-man-10} \alphab^\Ah \alphab^\Ah \Phi = 0\,,
\eeq
This constraint is manifestly invariant under the $osp(d,1|2)$ algebra. Note that $\Xi$ in \rf{02062026-man-35} should also satisfy the tracelessness constraint.

To complete our description we should present explicit expressions for the spin operators $B^\Ah$ and $M^{\Ah\Bh}$. The spin operator $M^{\Ah\Bh}$ for totally symmetric fields is well known and is given in \rf{21062026-man-50}, App.~C. The spin operator $B^\Ah$ is obtained by solving equations in \rf{02062026-man-92}.

\noinbf{Spin operator $B^\Ah$}. Solution for the spin operator $B^A$ we find is given by
\beq
\label{07062026-man-15} && B^\Ah = \bbf \alphab^\Ah + \alpha_\Tsm^\Ah \bar\bbf \,, \qquad \bbf := \zeta b_{N_\zeta}\,, \qquad \bar\bbf : = b_{N_\zeta}\zetab\,,
\nonumber\\
&& \hspace{1cm} \qquad  b_{N_\zeta} : = \NN_{N_\zeta} \BB_{N_{\zeta}}\,, \qquad \NN_{N_\zeta} := \Big(\frac{2s + d-3-N_\zeta}{2s +d-3 - 2 N_\zeta}\Big)^{1/2}\,,
\eeq
where the operators $\alpha_\Tsm^\Ah$ are given in \rf{21062026-man-55} of App.~C, while the operator $\BB_{N_\zeta}$ takes the form
\beq
\label{07062026-man-16} && \BB_{N_\zeta} = \big[(E_0 + s - 2 - N_\zeta)(E_0 - s - d + 2 + N_\zeta)\big]^{1/2}
\nonumber\\
&& \hspace{0.8cm}  = \big[ m^2 + N_\zeta (2s+d-4-N_\zeta) \big]^{1/2}\!, \hspace{1.8cm} \hbox{for massive field;}\qquad
\nonumber\\
&& \BB_{N_\zeta} =   \big[(t- N_\zeta)(2s + d - 4 - t - N_\zeta)\big]^{1/2}\,, \hspace{0.9cm} \hbox{for partially-massless field.}
\eeq
\! The remaining notation can be found in \rf{20062026-man-10} of App.~A. For the massive field in \rf{07062026-man-16}, the parameter $E_0$ stands for the lowest eigenvalue of the energy operator. Various helpful relations between the energy parameter $E_0$, mass parameter $m$, spin $s$, and depth $t$ may be found in App.~D. Before  proceeding, we introduce the definition  of

\noinbf{Classical unitarity}. In terms of the component fields, the kinetic terms for the field $\phi_0$ \rf{07062026-man-05} can schematically be presented as
{\small
\be \label{07062026-man-20}
\phi_0^\dagger (\Box+\partial_z^2) \phi_0 = \sum_{n} \frac{\varepsilon_n}{n!} \phi_0^{A_1\ldots A_n\dagger} (\Box+\partial_z^2) \phi_0^{A_1\ldots A_n}\,, \quad  \varepsilon_n-\hbox{real-valued}.
\ee
}
Recalling our choice $K=1$, we note that, if the BRST charge $Q$ and the measures $\varepsilon_n$ meet the following requirements
\beq
\label{07062026-man-25} \abf) \quad \Phi^\dagger Q^\dagger \Phi = \Phi^\dagger Q \Phi \,, \qquad \bbf) \quad \varepsilon_n > 0 \ \hbox{for all} \ n\,,
\eeq
then we refer our dynamical system to as classical unitary system.
The requirement \abf) tells us that the BRST-BV action is real-valued, while the requirement \bbf) implies that the kinetic terms of all field $\phi_0^{A_1\ldots A_n}$ appear the BRST-BV action with the correct sign. We make the following remarks.

\noinbf{i)} For the massive field, we find $\varepsilon_n=1$ for all $n$, while using solution \rf{07062026-man-16}, we verify, that the BRST charge satisfies requirement \abf). So, as expected, massive field is the classically unitary system. Needless to say that we assume $m^2>0$. Note also that requiring the eigenvalue of $B_{N_\zeta}$ to be real, we find the well-known unitarity  condition for a massive field, $E_0> s + d-2$.

\noinbf{ii)} For a partially-massless field, it is well known that the reality condition on the metric-like Lagrangian and requirement \bbf) cannot simultaneously be fulfilled. The same happens in the BRST-BV approach, where either requirement \abf) or \bbf) can be satisfied, but not both. We prefer to satisfy requirement \abf) and ignore \bbf). Using our solution in \rf{07062026-man-16}, we see that the requirement \abf) is fulfilled. Using then the commutators and the hermitian conjugation rules for the oscillator $\zeta$ given in \rf{20062026-man-05} in App.~A, we find $\varepsilon_n=(-)^{s-n}$  and hence the requirement \bbf) is not satisfied.

\noinbf{iii)} Our solution for the operator $B^\Ah$ is valid only on space of the fields that satisfy tracelessness constraint \rf{07062026-man-10}. This implies that our BRST charge is nilpotent only on the space of fields subject to the tracelessness constraint. By analogy with open string theory, it is preferable to get the $B^\Ah$ that satisfies the basic equations in the strong sense. Such solution is presented below.

\noinbf{iv)} For $M^{\Ah\Bh}$, hermiticity conditions \rf{19062026-man-37} with $K=1$ are  realized strongly, whereas for $B^\Ah$, these conditions are realized weakly. Namely, using the shorthand $X\approx Y$ in place of the equality $\Phi^\dagger X\Phi = \Phi^\dagger Y \Phi$, we note the following weak relations:
\be
B^{A\dagger} \approx B^A\,, \qquad B^{\eta\dagger} \approx B^\eta\,, \qquad B^{\rho\dagger} \approx - B^\rho\,.
\ee

\noinbf{v)} Plugging spin operators $B^\Ah$ \rf{07062026-man-15} and $M^{\Ah\Bh}$ into \rf{02062026-man-93}, we verify, as promised, that our solution for the operator $\CC_4$ leads to the correct eigenvalue of the 4th order Casimir operator of the $so(d,2)$ algebra. In App.~D, we recall eigenvalues of $\CC_4$ for all fields under investigation.

\noinbf{vi)}  For matching between our BRST-BV Lagrangian and the metric-like Lagrangian formulated in terms of the modified de Donder derivative, see App.~D. A comment on the derivation of solution \rf{07062026-man-15}, \rf{07062026-man-16} is given in App.~F.

\noinbf{Uniform constrained formulation}. Solution in \rf{07062026-man-15} is valid for an irreducible spin-$s$ massive/ partially-massless field. To obtain the uniform form of the solution that is valid for the finite or infinite tower of irreducible fields  \rf{03062026-man-38}, we use condition \rf{07062026-man-07} to remove the $s$ from the consideration. By doing so, we get the following uniform form of the solution:
\beq
\label{07062026-man-35} && B^\Ah = \bbf \alphab^\Ah + \alpha_\Tsm^\Ah \bar\bbf \,,\qquad \bbf = \zeta b_{N_\alphabfwh,N_\zeta}\,, \qquad \bar\bbf = b_{N_\alphabfwh,N_\zeta}\zetab\,,
\nonumber\\
&&   b_{N_\alphabfwh, N_\zeta} : = \NN_{N_\alphabfwh,N_\zeta} \BB_{N_\alphabfwh,N_{\zeta}}\,, \hspace{1cm} \NN_{N_\alphabfwh,N_\zeta}  = \Big(\frac{2N_\alphabfwh + N_\zeta + d-1}{2N_\alphabfwh +d-1}\Big)^{1/2}\,,
\eeq
where $ \alpha_\Tsm^\Ah$ are given in \rf{21062026-man-55} of App.~C and we use the notation
{\small
\beq
\label{07062026-man-36} && \BB_{N_\alphabfwh,N_\zeta}  = \big[(E_0 -1 + N_\alphabfwh)(E_0 -d+1- N_\alphabfwh)\big]^{1/2}
\nonumber\\
&& \hspace{1.3cm}  = \big[ m^2 + N_\zeta (2N_\alphabfwh +d-2+N_\zeta) \big]^{1/2}\!, \hspace{1.1cm}  \hbox{for tower of massive fields;}\qquad
\nonumber\\
&& \BB_{N_\alphabfwh,N_\zeta}  =\Big[\big(t-N_\zeta\big)\big(2N_\alphabfwh + N_\zeta + d-2 -t\big)\Big]^{1/2}\,, \hspace{0.3cm} \hbox{for tower of partially-massless fields}.\qquad
\eeq
}
\!The remaining notation can be found in \rf{20062026-man-10} of App.~A. We make the comments.

\noinbf{i)} The uniform solution is valid not only for a tower of irreducible fields but also for one irreducible field. Namely, using  condition \rf{07062026-man-07} in \rf{07062026-man-35}, we obtain back solution \rf{07062026-man-15}.

\noinbf{ii)} The eigenvalues of the operator $\BB_{N_\alphabfwh,N_\zeta}$ must be real in the entire space of fields \rf{03062026-man-38}.

\noinbf{iii)} For an irreducible field, $E_0$ and $t$ are $c$-numbers, whereas for a tower of fields, $E_0$ and $t$ are allowed to be operators constructed from the oscillators. Besides the above-mentioned restriction on the operator $\BB_{N_\alphabfwh,N_\zeta}$, the only restriction on operator-valued $E_0$ and $t$ is that they must (weakly) commute with all spin operators and the constraints. For example, $E_0$ and $t$ depending on $N_\alphabfwh+N_\zeta$ commute with all spin operators and weakly commute with constraint \rf{07062026-man-10}.

\noinbf{iv)}  In expressions for $\CC_2$, $\CC_4$ \rf{18062026-man-65}, one needs to make the substitution $s\rightarrow N_\alphabfwh + N_\zeta$, when treating equations \rf{02062026-man-92}, \rf{02062026-man-93}.

\subsection{\large Unconstrained formulation}

\noinbf{Field content}. The spin variables entering generating function $\Phi$ \rf{02062026-man-01} are fixed to be
\beq
\label{03062026-man-01ab} && \sbf = \alpha^\ah\,, \zeta\,, \chi\,,  \qquad \alpha^\ah = \alpha^a\,, \eta\,, \rho\,,\quad \chi^2 = 0 \,,
\eeq
where $\alpha^a$, $\zeta$, $\chi$ are Grassmann-even creation operators, while $\eta$, $\rho$ are Grassmann-odd creation operators. The oscillator $\alpha^\ah$ is a vector under the $osp(d-1,1|2)$ subalgebra. Expansion of $\phi$ and $\phi_*$ in the Grassmann-odd oscillators $\eta$, $\rho$ takes the form as in \rf{03062026-man-04}, where now the generating fields $\phi_0$, $\phi_\eta$, $\phi_\rho$, $\phi_{\eta\rho}$ depend on the oscillators $\alpha^a$, $\zeta$, $\chi$ and take the following form:
{\small
\beq
\label{03062026-man-10a-xx} && \hspace{-0.8cm}\phi_0 = \phi_{0\,0} + \chi \phi_{0\,1} \,, \quad \phi_\eta = \phi_{\eta\,0} + \chi \phi_{\eta\,1} \,, \quad \phi_\rho = \phi_{\rho\,0} + \chi \phi_{\rho\,1} \,, \quad \phi_{\eta\rho} = \phi_{\eta\rho\,0} + \chi \phi_{\eta\rho\,1} \,, \qquad
\nonumber\\
&& \phi_{0\, k} = \sum_{n=s-k-t\atop n\geq 0}^{s-k} \frac{\zeta^{s-k-n}\alpha^{a_1} \ldots \alpha^{a_n}}{n!\sqrt{(s-k-n)!}}  \phi_{0\, k}^{a_1\ldots a_n}\,,
\nonumber\\
&& \phi_{\eta,\rho\,k} = \sum_{n=s-1-k-t\atop n\geq 0}^{s-1-k} \frac{\zeta^{s-1-k-n}\alpha^{a_1} \ldots \alpha^{a_n}}{n!\sqrt{(s-1-k-n)!}}  \phi_{\eta,\rho\, k}^{a_1\ldots a_n}\,,
\nonumber\\
&& \phi_{\eta\rho\,k} = \sum_{n=s-2-k-t\atop n\geq 0}^{s-2-k} \frac{\zeta^{s-2-k-n}\alpha^{a_1} \ldots \alpha^{a_n}}{n!\sqrt{(s-2-k-n)!}}  \phi_{\eta\rho\, k}^{a_1\ldots a_n}\,,\hspace{2cm} k = 0,1\,,
\eeq
}
\!where in \rf{03062026-man-10a-xx} the shorthand $\phi_{\eta,\rho\,k}$ stands for the fields  $\phi_{\eta\,k}$ and $\phi_{\rho\,k}$ while the component fields $\phi^{a_1\ldots a_n}$ are scalar, vector, and totally symmetric tensor fields under the $so(d-1,1)$ subalgebra.  All tensor fields are traceful and there are no other constraints on the fields, meaning that we deal with the unconstrained formulation. It is easy to see that the field $\Phi$ satisfies the following homogeneity condition:
\beq
\label{03062026-man-15a} \big( N_\alphawh + N_\zeta + N_\chi- s) \Phi = 0\,,
\eeq
which tells us that $\Phi$ is degree-$s$ homogeneous polynomial in the oscillators $\alpha^\ah$, $\zeta$, $\chi$.

\noinbf{Solution for spin operators}. We discuss two equivalent formulations corresponding to the choice $K\ne 1$ and $K=1$.

\noindent {\it Case $K\ne 1$}. For this case, the operator $K$ and spin operators  $M^{\ah\bh}$, $M^{z\ah}$ take the forms given in relations \rf{23062026-man-01}, \rf{23062026-man-05} of App.~E, while the operator $B^\Ah$ can be presented as
\be
B^\Ah = \bbf \alphab_\chi^\Ah + \alpha_{\Tsm\chi}^\Ah \bar\bbf\,,
\ee
where the operators $\alphab_\chi^\Ah$, $\alpha_{\Tsm\chi}^\Ah$ are defined in \rf{23062026-man-10}. The explicit form of the operators $\bbf$ and $\bar\bbf$ depends on whether we describe an irreducible field or a tower of irreducible fields. We note that

\noinbf{i)} For the irreducible spin-$s$ field that satisfies constraint \rf{03062026-man-15a}, the operators $\bbf$ and $\bar\bbf$ take the form given in \rf{07062026-man-15}, \rf{07062026-man-16}.

\noinbf{ii)} For tower of fields \rf{03062026-man-38}, the operators $\bbf$ and $\bar\bbf$ take largely uniform form \rf{07062026-man-35}, \rf{07062026-man-36}. The only required modification is that we should make the replacement $N_\alphabfwh\rightarrow N_\alphawh + N_\chi$ in \rf{07062026-man-35}, \rf{07062026-man-36}.  Note also that we should make the replacement $s\rightarrow N_\alphawh + N_\chi + N_\zeta$, when dealing with equations  \rf{02062026-man-92}, \rf{02062026-man-93}.

\noinbf{iii)} The relations  $Q^2=0$, $(KQ)^\dagger= KQ$, and hermiticity conditions \rf{19062026-man-37} for the spin operators hold strongly. $B^\Ah$, $M^{\Ah\Bh}$ satisfy equations \rf{02062026-man-92} in the strong sense.

\noindent {\it Case $K = 1$}. For this case, the spin operators  $M^{\ah\bh}$, $M^{z\ah}$ take the forms given in relations \rf{23062026-man-30}, \rf{23062026-man-35} of App.~E, while the operator $B^\Ah$ can be presented as
\be
B^\Ah = \bbf \alphab_\Ksm^\Ah + \alpha_\Ksm^\Ah \bar\bbf\,,
\ee
where the operators $\alphab_\Ksm^\Ah$, $\alpha_\Ksm^\Ah$ are defined in \rf{23062026-man-35}. The explicit form of the operators $\bbf$ and $\bar\bbf$ depends on whether we describe an irreducible field or a tower of irreducible fields. We note that

\noinbf{i)} For the irreducible spin-$s$ field that satisfies  constraint \rf{03062026-man-15a}, the operators $\bbf$ and $\bar\bbf$ take the same form as the ones given in \rf{07062026-man-15}, \rf{07062026-man-16}.

\noinbf{ii)} For tower of fields \rf{03062026-man-38}, the operators $\bbf$ and $\bar\bbf$ take largely the uniform representation given in \rf{07062026-man-35}, \rf{07062026-man-36}. The only required modification is that we should make the replacement $N_\alphabfwh\rightarrow N_\alphawh + N_\chi$ in \rf{07062026-man-35}, \rf{07062026-man-36}. Note also that we should make the replacement $s\rightarrow N_\alphawh + N_\chi + N_\zeta$  when treating equations  \rf{02062026-man-92}, \rf{02062026-man-93}.

\noinbf{iii)} The relations  $Q^2=0$, $Q^\dagger= Q$, and hermiticity conditions \rf{19062026-man-37} for the spin operators with $K=1$ hold strongly. $B^\Ah$, $M^{\Ah\Bh}$ satisfy equations \rf{02062026-man-92} in the strong sense.

\newsection{ \large Continuous-spin field }\label{sec-05}

The Lagrangian description of a continuous-spin field (CSF) proposed in Refs.\cite{Schuster:2013pta,Najafizadeh:2015uxa} has stimulated  considerable interest in studying its various aspects. For review, see Refs.\cite{Bekaert:2006py,Brink:2002zx,Bekaert:2017khg}. We now provide an inevitably incomplete list of references to recent studies.

For the study of an interacting CSF, the use of a worldline approach may be found in Refs.\cite{Schuster:2023xqa,Schuster:2023jgc,Kundu:2025mzm}, while the use of a field-theoretical approach is available in Ref.\cite{Metsaev:2017cuz,Bekaert:2017xin,Rivelles:2018tpt,Metsaev:2018moa, Metsaev:2025qkr,Metsaev:2025wcv}. Various interesting applications of CSF to physics are considered in Refs.\cite{Schuster:2024wjc,Kundu:2026zmb}.
The Lagrangian metric-like and frame-like formulations of CSF in AdS were discussed in Refs.\cite{Metsaev:2016lhs,Metsaev:2017ytk,Khabarov:2017lth}, while the light-cone gauge formulation of CSF was studied in Refs.\cite{Metsaev:2019opn,Metsaev:2025nbm}. Mixed-symmetry CSF in AdS and flat spaces were discussed in Refs.\cite{Khabarov:2017lth,Metsaev:2017myp,Alkalaev:2017hvj, Metsaev:2021zdg,Golubtsova:2025eei}. Within the BRST approach, the study of CSF in flat space is presented in Refs.\cite{Bengtsson:2013vra,Metsaev:2018lth,Buchbinder:2018yoo, Buchbinder:2020nxn,Burdik:2019tzg}.
In $AdS_4$, BRST studies of CSF may be found in Refs.\cite{Buchbinder:2024jpt,Buchbinder:2024vli}, while a continuous-spin (super)particle was considered in Refs.\cite{Buchbinder:2024hea,Buchbinder:2025gsr}.
Wigner constraints for CSF in AdS were investigated in Ref.\cite{Astrakhantsev:2026mby}, while the  study of CSF in flat space via the universal model of a spinning particle may be found in Ref.\cite{Lyakhovich:1996we}. Supersymmetric CSF was investigated in Refs.\cite{Buchbinder:2019kuh,Najafizadeh:2019mun}. For a more extensive list of references to the recent and earlier literature, see Ref.\cite{Basile:2026ntj}.

We now describe the BRST-BV formulation for CSF in $AdS_{d+1}$, $d\geq 3$.

\subsection{\large Constrained formulation for irreducible CSF}

\noinbf{Field content}. The kernel $K$ is fixed to be $K=1$, while the spin variables in \rf{02062026-man-01} are fixed to be
\beq
\label{08062026-man-01} && \sbf := \alpha^\Ah\,, \upsilon\,, \qquad \qquad \alpha^\Ah: = \alpha^A\,, \eta\,, \rho\,, \qquad
\eeq
where $\alpha^A$, $\upsilon$ are Grassmann-even creation operators, while  $\eta$, $\rho$ are Grassmann-odd creation operators.  Expansion of $\phi$ and $\phi_*$ in $\eta$, $\rho$ takes the form
\beq
\label{08062026-man-03} && \phi = \phi_0 + \eta \phi_\eta + \rho \phi_\rho + \rho\eta \phi_{\eta\rho}\,, \qquad  \phi_* = \phi_{*0} + \eta \phi_{*\eta} + \rho \phi_{*\rho} + \rho\eta \phi_{*\eta\rho}\,,
\eeq
where the fields $\phi_0$, $\phi_\eta$, $\phi_\rho$, $\phi_{\eta\rho}$ depend only on the oscillators $\alpha^A$, $\upsilon$  and take the following form in terms of component fields:
{\small
\beq
\label{08062026-man-05} &&  \phi_0 = \sum_{n=0}^\infty \frac{\upsilon^n}{n!\sqrt{n!}} \alpha^{A_1} \ldots \alpha^{A_n} \phi_0^{A_1\ldots A_n}\,,
\nonumber\\
&&   \phi_{\eta,\rho} = \sum_{n=0}^\infty \frac{\upsilon^{n+1}}{n!\sqrt{(n+1)!}}  \alpha^{A_1} \ldots \alpha^{A_n} \phi_{\eta,\rho}^{A_1\ldots A_n}\,,
\nonumber\\
&& \phi_{\eta\rho} = \sum_{n=0 }^\infty \frac{\upsilon^{n+2}}{n!\sqrt{(n+2)!}} \alpha^{A_1} \ldots \alpha^{A_n} \phi_{\eta\rho}^{A_1\ldots A_n}\,,
\eeq
}
\!while the fields $\phi^{A_1\ldots A_n}$ appearing in \rf{08062026-man-05}, are scalar, vector, and totally symmetric tensor fields under the $so(d,1)$ Lorentz algebra. It is easy to see that the field $\Phi$ satisfies the following homogeneity condition
\beq
\label{08062026-man-07}  && \big( N_\alphabfwh - N_\upsilon) \Phi = 0\,.
\eeq

To obtain solution for an irreducible CSF, we impose the tracelessness constraint
\beq
\label{08062026-man-10} \alphab^\Ah \alphab^\Ah \Phi = 0\,,
\eeq
and note that $\Xi$ entering gauge transformation \rf{02062026-man-35} should also satisfy the tracelessness constraint.

\noinbf{Spin operators $B^\Ah$ and $M^{\Ah\Bh}$}. As usually for totally symmetric field, the spin operator $M^{\Ah\Bh}$ takes the form given in \rf{21062026-man-50}. The solution for the spin operator $B^A$ we obtain is given by
\beq
\label{08062026-man-25} && B^\Ah = \bbf \alphab^\Ah + \alpha_\Tsm^\Ah \bar\bbf \,,\qquad \bbf =  f_\upsilon  \NN_\upsilon \upsilonb\,, \qquad \bar\bbf = \upsilon \NN_\upsilon \fb_\upsilon\,,\qquad f_\upsilon  \fb_\upsilon = F_\upsilon\,,
\nonumber\\
&& \hspace{1cm}  F_\upsilon : = \big((N_\upsilon + \frac{d-2}{2})^2 - p^2\big) \big((N_\upsilon + \frac{d-2}{2})^2 - q^2\big)
\nonumber\\
&& \hspace{1.8cm} =\,\, l_p l_{-p} l_q l_{-q}\,, \qquad l_X: = N_\upsilon + \frac{d-2}{2} + X
\nonumber\\
&& \hspace{1cm} \NN_{N_\upsilon} : = \big[(N_\upsilon+1)(2N_\upsilon+d-1)\big]^{-1/2}\,,
\eeq
where $ \alpha_\Tsm^\Ah$ are given in \rf{21062026-man-55}, while $p$ and $q$ are labels which we prefer to use in place of the energy parameter $E_0$ and the spin parameter $s$ (see \rf{18062026-man-75} in App.~D). In general, the labels $p$ and $q$ are complex-valued and have been used in Ref.\cite{Metsaev:2019opn} to classify continuous-spin fields. The quantities $f_\upsilon$, $\fb_\upsilon$, where $f_\upsilon^\dagger = \fb_\upsilon$, depend only on $N_\upsilon$, $p$, and $q$. The explicit expressions for $f_\upsilon$ and $\fb_\upsilon$, corresponding to various values of the labels $p$ and $q$ are given in (3.20)-(3.24) of Ref.\cite{Metsaev:2019opn}.

\noinbf{i)} For CSF, the classical unitarity requirement \bbf) \rf{07062026-man-25} is automatically fulfilled, while the classical unitarity requirement \abf) \rf{07062026-man-25} amounts to the requirement that the eigenvalues of the operator $F_\upsilon$ be real and positive. In turn, the latter requirement amounts to restrictions on the allowed values of the labels $p$ and $q$. These restrictions completely coincide with those  obtained within the light-cone gauge approach in Ref.\cite{Metsaev:2019opn} (see eq.(3.13) therein and the additional clarifying remark in Sec.~2 of Ref.\cite{Metsaev:2025nbm}). This implies that the classification of classically unitary CSF within the BRST-BV approach we discuss here completely coincides with the one obtained within the light-cone gauge approach in Refs.\cite{Metsaev:2019opn,Metsaev:2025nbm}. In other words, our study here provides the BRST-BV description of CSF in $AdS_{d+1}$, $d\geq3$, for all allowed values of the labels $p$, $q$ described in Refs.\cite{Metsaev:2019opn,Metsaev:2025nbm}.

\noinbf{ii)} Plugging spin operators $B^\Ah$ \rf{08062026-man-25} and $M^{\Ah\Bh}$ \rf{21062026-man-50} into \rf{02062026-man-93}, we verify that our solution for the operator $\CC_4$ leads to the correct eigenvalue of the 4th order Casimir operator of the $so(d,2)$ algebra corresponding to CSF (see \rf{18062026-man-75} in App.~D).

\subsection{\large Unconstrained formulation for irreducible CSF}

\noinbf{Field content}. The spin variables entering generating function $\Phi$ \rf{02062026-man-01} are fixed to be
\beq
\label{08062026-man-50}  && \sbf = \alpha^\ah\,, \upsilon\,, \chi\,,  \qquad \alpha^\ah = \alpha^a\,, \eta\,, \rho\,,\quad \chi^2 = 0 \,,
\eeq
where $\alpha^a$, $\upsilon$, $\chi$ are Grassmann-even creation operators, while $\eta$, $\rho$ are Grassmann-odd creation operators. The oscillator $\alpha^\ah$ is a vector under the $osp(d-1,1|2)$ subalgebra. Expansion of $\phi$ and $\phi_*$ in the Grassmann-odd oscillators $\eta$, $\rho$ takes the form as in \rf{08062026-man-03}, where now the generating fields $\phi_0$, $\phi_\eta$, $\phi_\rho$, $\phi_{\eta\rho}$ depend on the oscillators $\alpha^a$, $\upsilon$, $\chi$ and take the following form
{\small
\beq
\label{08062026-man-55}  && \hspace{-0.8cm}\phi_0 = \phi_{0\,0} + \chi \phi_{0\,1} \,, \quad \phi_\eta = \phi_{\eta\,0} + \chi \phi_{\eta\,1} \,, \quad \phi_\rho = \phi_{\rho\,0} + \chi \phi_{\rho\,1} \,, \quad \phi_{\eta\rho} = \phi_{\eta\rho\,0} + \chi \phi_{\eta\rho\,1} \,, \qquad
\nonumber\\
&& \phi_{0\, k} = \sum_{n= 0}^\infty \frac{\upsilon^{n+k}\alpha^{a_1} \ldots \alpha^{a_n}}{n!\sqrt{(n+k)!}}  \phi_{0\, k}^{a_1\ldots a_n}\,,
\nonumber\\
&& \phi_{\eta,\rho\,k} = \sum_{n=0}^\infty \frac{\upsilon^{n+1+k}\alpha^{a_1} \ldots \alpha^{a_n}}{n!\sqrt{(n+1+k)!}}  \phi_{\eta,\rho\, k}^{a_1\ldots a_n}\,,
\nonumber\\
&& \phi_{\eta\rho\,k} = \sum_{n= 0}^\infty \frac{\upsilon^{n+2+k}\alpha^{a_1} \ldots \alpha^{a_n}}{n!\sqrt{(n+2+k)!}}  \phi_{\eta\rho\, k}^{a_1\ldots a_n}\,,\hspace{2cm} k = 0,1\,,
\eeq
}
\!where, in \rf{08062026-man-55}, the shortcut $\phi_{\eta,\rho\,k}$ stands for the fields  $\phi_{\eta\,k}$ and $\phi_{\rho\,k}$. The component fields $\phi^{a_1\ldots a_n}$ in \rf{08062026-man-55} are scalar, vector, and totally symmetric tensor fields under the $so(d-1,1)$ subalgebra. All tensor fields are traceful and there are no other constraints on the fields, meaning that we deal with the unconstrained formulation. It is easy to see that the field $\Phi$ satisfies the following homogeneity condition:
\beq
\label{08062026-man-60} && \big( N_\alphawh + N_\chi- N_\upsilon) \Phi = 0\,.
\eeq

\noinbf{Solution for spin operators}. We discuss two equivalent formulations corresponding to the cases $K\ne 1$ and $K=1$.

\noindent {\it Case $K\ne 1$}. For this case, the operator $K$ and spin operators  $M^{\ah\bh}$, $M^{z\ah}$ take the forms given in relations \rf{23062026-man-01}, \rf{23062026-man-05} of App.~E, while the operator $B^\Ah$ can be presented as
\be
B^\Ah = \bbf \alphab_\chi^\Ah + \alpha_{\Tsm\chi}^\Ah \bar\bbf\,,
\ee
where the operators $\alphab_\chi^\Ah$, $\alpha_{\Tsm\chi}^\Ah$ are defined in \rf{23062026-man-10}. The operators $\bbf$ and $\bar\bbf$ take the same form as the ones given in \rf{08062026-man-25}. The relations  $Q^2=0$, $(KQ)^\dagger= KQ$, and hermiticity conditions \rf{19062026-man-37} for the spin operators hold strongly. $B^\Ah$, $M^{\Ah\Bh}$ satisfy equations \rf{02062026-man-92} in the strong sense.

\noindent {\it Case $K = 1$}. For this case, the spin operators  $M^{\ah\bh}$, $M^{z\ah}$ take the forms given in relations \rf{23062026-man-30}, \rf{23062026-man-35} of App.~E, while the operator $B^\Ah$ can be presented as
\be
B^\Ah = \bbf \alphab_\Ksm^\Ah + \alpha_\Ksm^\Ah \bar\bbf\,,
\ee
where the operators $\alphab_\Ksm^\Ah$, $\alpha_\Ksm^\Ah$ are defined in \rf{23062026-man-35}. The operators $\bbf$ and $\bar\bbf$ take the same form as the ones given in \rf{08062026-man-25}.  The relations  $Q^2=0$, $Q^\dagger= Q$, and hermiticity conditions \rf{19062026-man-37} for the spin operators, where $K=1$, are satisfied  strongly. $B^\Ah$, $M^{\Ah\Bh}$ satisfy equations \rf{02062026-man-92} in the strong sense.

\newsection{ Conclusions } \label{sec-06}

Using Poincar\'e parametrization of AdS, we developed a general BRST-BV Lagrangian
formulation for AdS fields. We expressed the BRST charge in terms of the AdS mass operator and the spin operators. We found closed and relatively simple equations for the spin operators. Importantly, our approach is applicable on an equal footing to both irreducible fields and towers of fields. Also, our framework can be used for AdS fields having arbitrary type of symmetry. For illustrative purposes, we applied our method to the study of totally symmetric fields. We considered massless, massive, partially-massless fields with arbitrary integer spin and the continuous-spin field. We discussed both irreducible fields and towers of irreducible fields. We developed the constrained and unconstrained formulations. Perhaps the most attractive feature of our approach is that it reduces the problem of finding a BRST-BV description of fields in AdS to a purely algebraic problem: finding the spin operators by solving defining equations for them. Our approach might have the following interesting applications and generalizations.

The BRST approach is a powerful method for studying interacting fields. Applications of this approach to the study of massless and massive fields may be found, e.g., in Refs.\cite{Bekaert:2004dz,Metsaev:2012uy,Dempster:2012vw,Polyakov:2015usr, Grigoriev:2020lzu,Buchbinder:2021xbk,Reshetnyak:2023lqj}. The intriguing BRST model of higher-spin gravity with supersymmetry in any dimension is discussed in Ref.\cite{Vasiliev:2025hfh}, while the interesting application of the BRST method to the study of spinning black holes may be found in Refs.\cite{Skvortsov:2023jbn,Lang:2026iaz}. Our BRST-BV formulation turns out to be very algebraic in nature. Therefore, we expect that the applications of our approach to interacting fields in AdS is a promising direction for future research.

Supersymmetric theories of higher-spin fields remain a focus of current research.
Cubic interactions for $\NN=1$ and $\NN=4N$ higher-spin supermultiplets in flat space were considered in Refs.\cite{Metsaev:2019dqt,Metsaev:2019aig}, while the AdS counterparts of the vertices in Ref.\cite{Metsaev:2019dqt} were found in Ref.\cite{Khabarov:2020deh}. Recent progress in studying interactions of $\NN=2$ higher-spin supermultiplets by using harmonic superspace may be found in Refs.\cite{Buchbinder:2022vra,Buchbinder:2024pjm,Zaigraev:2026jvu,Zaigraev:2026xzr}. Application of our approach to the study of supersymmetric higher-spin theories could be of some interest.

\setcounter{section}{0}
\appendix{ \large Notation and conventions  }\label{app-a}

Vector indices of the Lorentz algebra  $so(d,1)$ run over $A,B,C,E=0,1,\ldots, d$, while vector indices of the Lorentz algebra  $so(d-1,1)$ run over $a,b,c,e=0,1,\ldots ,d-1$.  Assuming the identification $X^z:= X^d$, a vector of the $so(d,1)$ algebra is then decomposed as $X^A = (X^a,X^z)$. In scalar products, we omit the explicit flat metrics: $X^AY^A := \eta_{AB}X^A Y^B$, $X^aY^a := \eta_{ab}X^a Y^b$, and hence $X^AY^A = X^a Y^a + X^z Y^z$.
We use mostly positive flat metrics: $\eta^{AB}=(-,+,\ldots,+)$,  $\eta^{ab}=(-,+,\ldots,+)$.
The Poincar\'e parametrization of $AdS_{d+1}$ space is given by:
\be
ds^2 = \frac{1}{z^2}(dx^a dx^a + dz dz)\,.
\ee

For the space-time derivatives, we use the notation: $\partial^a := \eta^{ab}\partial/\partial x^b$, $\partial_z := \partial/\partial z$, $\partial^A := (\partial^a,\partial_z)$. The left derivative for the Grassmann coordinate $\theta$ is denoted by $\partial_\theta$. The integral over $\theta$ is normalized as $\int d\theta \theta =1$. Hermitian conjugation rules for the coordinates and the derivatives are defined as $(x^A,\theta)^\dagger = (x^A,\theta)$, $(\partial^A, \partial_\theta)^\dagger  = (-\partial^A,\partial_\theta)$. For the product of two operators $A$ and $B$ with arbitrary Grassmann parities, Hermitian conjugation is defined according to the rule $(AB)^\dagger = B^\dagger A^\dagger$.

We use the Grassmann-even creation operators $\alpha^A$, $\zeta$, $\upsilon$, $\chi$ and the Grassmann-odd creation operators $\eta$, $\rho$. The corresponding annihilation  operators are denoted as $\alphab^A$, $\zetab$, $\upsilonb$, $\chib$ and $\rhob$, $\etab$. We collectively refer to these operators as oscillators. The (anti)commutators, the definition of the vacuum $|0\rangle$, and the Hermitian conjugation rules are given by:
{\small
\beq
\label{20062026-man-05}  && [\alphab^A,\alpha^B]\! =\! \eta^{AB},  \hspace{0.5cm}  [\zetab,\zeta]= \epsilon, \hspace{0.5cm} [\upsilonb,\upsilon]=1\,, \hspace{0.5cm} \{\chib,\chi\} = 1\,, \hspace{0.5cm} \{\rhob,\eta\}=1\,, \hspace{0.5cm} \{\etab,\rho\} =1\,, \qquad
\nonumber\\
&& \alphab^A |0\rangle = 0\,, \hspace{1.2cm} \zetab |0\rangle = 0\,,  \hspace{0.5cm} \upsilonb |0\rangle = 0\,,  \hspace{0.7cm} \chib |0\rangle=0\,,\hspace{0.8cm} \etab |0\rangle=0\,,\hspace{0.8cm} \rhob |0\rangle=0\,,
\nonumber\\
&& \alpha^{A \dagger} = \alphab^A\,, \hspace{1.3cm} \zeta^\dagger = \zetab\,, \hspace{0.8cm} \upsilon^\dagger = \upsilonb\,, \hspace{1cm} \chi^\dagger = \chib\,, \hspace{1cm} \eta^\dagger = \etab\,,\hspace{1cm}\rho^\dagger = \rhob\,,
\nonumber\\
&& \epsilon = 1  \hbox{ for massive fields}; \hspace{1cm} \epsilon = -1  \hbox{ for partially-massless fields}.
\eeq
}
\!Note also that $\chi^2=\eta^2=\rho^2=0$ and hence $\chib^2=\etab^2=\rhob^2=0$. The oscillators $\alpha^A$, $\alphab^A$  are vectors under the $so(d,1)$ Lorentz algebra, while all the remaining oscillators are scalars under this algebra.

Throughout this paper the following notation for the products of the derivatives and the oscillators is used:
\beq
\label{20062026-man-10} && N_\alphabfwh := \alpha^\Ah \alphab^\Ah\,, \hspace{0.5cm} N_\alphawh : = \alpha^\ah \alphab^\ah\,, \hspace{0.5cm} N_\alphabf : = \alpha^A \alphab^A  \,, \hspace{0.5cm} N_\alpha := \alpha^a \alphab^a  \,, \hspace{0.3cm} N_{\alpha^z} :=\alpha^z\alphab^z\,,
\nonumber\\
&& N_\zeta : = \epsilon \zeta \zetab \,,\hspace{0.9cm} N_\upsilon: = \upsilon\upsilonb\,, \hspace{1.1cm}  N_\eta := \eta \rhob\,, \hspace{1.1cm}  N_\rho := \rho \etab\,, \hspace{0.9cm}  N_\chi := \chi \chib\,,\qquad
\eeq
where $\epsilon$ is given in \rf{20062026-man-05} and we note the relations $N_\alphabfwh=N_\alphawh + N_{\alpha^z}$, $N_\alphabf=N_\alpha + N_{\alpha^z}$.

Throughout this paper, we omit the bra-ket notation. For example, a quadratic form $\Phi^\dagger X \Phi$ appearing in a Lagrangian, where an operator $X$ is composed of oscillators, is defined as
\beq
\Phi^\dagger X \Phi : = \langle \Phi|X|\Phi\rangle\,, \quad |\Phi\rangle: = \Phi|0\rangle\,, \quad \langle \Phi|:= (|\Phi\rangle)^\dagger\,.
\eeq
Also, we use the shorthand $X\Phi$ in place of the expression $X|\Phi\rangle$.

\noinbf{Ghost numbers and hermiticity conditions}. On the space of $\Phi$, the internal Faddeev-Popov ghost operator $N_\FPsm^\intrm$  is realized as
\be \label{20062026-man-50}
N_\FPsm^\intrm = \theta\partial_\theta + N_\eta - N_\rho\,.
\ee
The ghost numbers of the oscillators, $\theta$, $\partial_\theta$, and the vacuum  are then given by
{\small
\beq
\label{20062026-man-55} && \gh(\alpha^A,\alphab^A,\zeta,\zetab,\upsilon,\upsilonb,\chi,\chib) =0\,,\quad \gh(\theta, \eta, \rhob) = 1\,, \quad \gh(\partial_\theta, \etab, \rho) = -1\,,\quad \gh|0\rangle=0\,, \qquad
\eeq
}
while the ghost number of fields and antifields are given by
\beq
&& \gh(\phi_0)=0\,, \hspace{1cm} \gh(\phi_\eta)=-1\,, \hspace{1cm} \gh(\phi_\rho)=1\,, \hspace{1cm} \gh(\phi_{\eta\rho})=0\,,
\nonumber\\
&& \gh(\phi_{*0})=-1\,, \hspace{0.5cm} \gh(\phi_{*\eta})=-2\,, \hspace{0.7cm} \gh(\phi_{*\rho})=0\,,  \hspace{0.9cm} \gh(\phi_{*\eta\rho}) = -1\,.\qquad
\eeq
Recall the relation for the ghost numbers of fields and antifields: $\gh(\hbox{field})+\gh(\hbox{antifield})=-1$.

Throughout this paper, we use a basis of complex-valued fields and antifields. For the reader's convenience, we present hermiticity conditions which are required for the transition to a real-valued basis. Using the shorthand $\phi^{A(n)}$ for $\phi^{A_1\ldots A_n}$, we note the following hermiticity conditions:
\beq
&& \phi_0^{A(n)\dagger} = \phi_0^{A(n)}\,, \hspace{0.6cm} \phi_\eta^{A(n)\dagger} = - \phi_\eta^{A(n)}\,, \hspace{0.3cm} \phi_\rho^{A(n)\dagger} = \phi_\rho^{A(n)}\,, \hspace{0.3cm} \phi_{\eta\rho}^{A(n)\dagger} =  \phi_{\eta\rho}^{A(n)}\,.
\nonumber\\
&& \phi_{*0}^{A(n)\dagger} = - \phi_{*0}^{A(n)}\,, \hspace{0.3cm} \phi_{*\eta}^{A(n)\dagger} = - \phi_{*\eta}^{A(n)}\,, \hspace{0.3cm} \phi_{*\rho}^{A(n)\dagger} = \phi_{*\rho}^{A(n)}\,, \hspace{0.3cm} \phi_{*\eta\rho}^{A(n)\dagger} =  -\phi_{*\eta\rho}^{A(n)}\,.
\eeq

\noinbf{Terminology in the literature}. For brevity, we refer to the component fields in $\phi$ \rf{03062026-man-04} as fields, whereas the component fields in $\phi_*$ \rf{03062026-man-04} we refer to as antifields. The more popular terminology is as follows:
\ibf) $\phi_0$, $\phi_{\eta\rho}$, $\phi_{*\rho}$ appearing in $\phi$ and $\phi_*$ have $\gh=0$ and they are referred to as fields; \iibf) $\phi_\rho$ appearing in $\phi$ has $\gh=1$ and it is referred to as a ghost; \iiibf) $\phi_\eta$ appearing in $\phi$ and $\phi_{0*}$, $\phi_{*\eta\rho}$ appearing in $\phi_*$ have negative ghost numbers and they are referred to as antifields. Upon using the Siegel gauge $\phi_*=0$, $\phi_\rho$ and $\phi_\eta$ are realized as the Faddeev-Popov ghost and anti-ghost, respectively. Note  that the fields $\phi_0$, $\phi_{\eta\rho}$, and $\phi_{*\rho}$ constitute a so-called triplet.

\appendix{ \large $osp(d,1|2)$ algebra  and its $osp(d-1,1|2)$ subalgebra}\label{app-b}

\noinbf{\bf $so(d,1)\oplus sp_{\eta\rho}(2)$-basis}. In the basis under consideration, the generators of the $osp(d,1|2)$ algebra can be presented as:
\beq
\label{19062026-man-24}  \underbrace{ M^{AB}}_{so(d,1)} \quad \underbrace{ M^{\eta\rho},M^{\eta\eta},M^{\rho\rho} }_{sp(2)}\quad \underbrace{M^{\eta A},M^{\rho A}}_{{{\scriptstyle \rm supercharges}}}\,.
\eeq
The (anti)commutators are given by
{\small
\beq
\label{19062026-man-25} && \hbox{\bf commutators of $so(d,1)$ and $sp(2)$}:
\nonumber\\
&& [M^{AB},M^{CE}] = \eta^{BC} M^{AE} + 3 \hbox{ terms},
\nonumber\\
&& [M^{\eta\rho},M^{\eta \eta}] = - 2 M^{\eta\eta}\,,\quad [M^{\eta\rho},M^{\rho \rho}] =  2 M^{\rho\rho}\,,\quad [M^{\eta\eta},M^{\rho\rho}] = 4 M^{\eta\rho}\,,\qquad
\nonumber\\
&& \hbox{\bf cross commutators between $so(d,1)$, $sp(2)$ and supercharges}:
\nonumber\\
&& [M^{AB},M^{\eta C}] = \eta^{BC} M^{\eta A} - \eta^{AC} M^{\eta B} \,,\qquad [M^{AB}, M^{\rho C}] = \eta^{BC} M^{\rho A} - \eta^{AC} M^{\rho B}\,,
\nonumber\\
&& [M^{\eta\rho},M^{\eta A}] = -M^{\eta A}\,,\qquad [M^{\rho\rho},M^{\eta A}] = - 2M^{\rho A}\,,
\nonumber\\
&& [M^{\eta\rho},M^{\rho A}] =  M^{\rho A}\,,\qquad \ \ \ [M^{\eta\eta},M^{\rho A}] = 2 M^{\eta A}\,,
\nonumber\\
&& \hbox{\bf anti-commutators between supercharges}:
\nonumber\\
&& \{M^{\eta A}, M^{\rho B}\} = - \eta^{AB} M^{\eta\rho}  - M^{AB}\,,
\nonumber\\
&& \{ M^{\eta A},M^{\eta B}\} = - \eta^{AB} M^{\eta\eta}\,,\qquad \{ M^{\rho A},M^{\rho B}\} =  - \eta^{AB} M^{\rho\rho}\,.\qquad
\eeq
}
We note $M^{AB}=-M^{BA}$, $M^{\eta\rho}=M^{\rho\eta}$, $M^{\eta A}=-M^{A\eta}$, $M^{\rho A}=-M^{A\rho}$.

Let us denote the components of the $osp(d,1|2)$ algebra vector as $X^A$, $X^\eta$, $X^\rho$, where $X^A$ is a Grassmann-even vector under the $so(d,1)$ Lorentz algebra, while $X^\eta$, $X^\rho$ are Grassmann-odd scalars under this algebra. Commutators between the generators of the $osp(d,1|2)$ algebra and the above-mentioned components of the vector take then the form:
{\small
\beq
\label{19062026-man-30} && \hspace{-0.7cm} [M^{AB},X^C] = \eta^{BC} X^A - \eta^{AC} X^B \,,
\nonumber\\
&&  \hspace{-0.7cm} [M^{\eta\rho}, X^\eta] = -  X^\eta\,, \qquad [M^{\rho\rho}, X^\eta] = - 2 X^\rho\,, \qquad [M^{\eta\rho}, X^\rho] =  X^\rho\,, \qquad \ [M^{\eta\eta}, X^\rho] = 2 X^\eta\,,\qquad
\nonumber\\
&& \hspace{-0.7cm} [M^{\eta A}, X^B] = \eta^{AB} X^\eta,\quad \{M^{\eta A}, X^\rho\} = -  X^A\,,\quad [M^{\rho A}, X^B] = \eta^{AB} X^\rho,\quad \{M^{\rho A}, X^\eta\} =  X^A\,.\qquad
\eeq
}
An invariant scalar product $(X,Y)$ of two vectors of the $osp(d,1|2)$ algebra is given by:
\be \label{19062026-man-35}
X\cdot Y = X^A Y^A -  X^\eta Y^\rho + X^\rho Y^\eta\,, \qquad X^A Y^A: = \eta_{AB} X^A Y^B\,.
\ee

The spin operators and the operators $B^A$, $B^\eta$, and $B^\rho$ satisfy the following hermiticity conditions:
{\small
\beq
\label{19062026-man-37}  && (KM^{AB})^\dagger = - KM^{AB}\,, \quad (KM^{\eta\rho})^\dagger = KM^{\eta\rho}\,, \quad  (KM^{\eta\eta})^\dagger = KM^{\eta\eta}\,, \quad (KM^{\rho\rho})^\dagger =  KM^{\rho\rho}\,,
\nonumber\\
&& (KM^{\eta A})^\dagger = - KM^{\eta A}\,, \ \quad (KM^{\rho A})^\dagger = KM^{\rho A}\,,
\nonumber\\
&& (KB^A)^\dagger =  KB^A\,, \hspace{1.4cm} (KB^\eta)^\dagger = KB^\eta\,,\hspace{1cm} (KB^\rho)^\dagger = - KB^\rho\,,
\eeq
}
\!where $K$ is a kernel appearing in the BRST-BV Lagrangian \rf{02062026-man-30}. Depending on the formulation, hermiticity conditions \rf{19062026-man-37} are realized either strongly or weakly.

\noindent{\bf $so(d,1)\oplus sp(2)$-basis}. In this basis, the $sp(2)$ algebra generators $M^{\isf\jsf}$ and supercharges  $M^{\isf A}$, where $\isf,\jsf,\ksf,\lsf=1,2$, are identified with those in \rf{19062026-man-24} as:
{\small
\beq
&& M^{11} = M^{\eta\eta}\,,\quad M^{12} =M^{21} = M^{\eta\rho}\,,\quad M^{22} = M^{\rho\rho}\,,\quad M^{1A} = M^{\eta A}\,,\quad M^{2A} = M^{\rho A}\,,\qquad
\eeq
}
\!and hence, in the basis under consideration, the generators of the $osp(d,1|2)$ algebra \rf{19062026-man-24} can be presented as:
\beq
\label{19062026-man-36} \underbrace{ M^{AB}}_{so(d,1)} \quad \underbrace{ M^{\isf\jsf} }_{sp(2)}\quad \underbrace{M^{\isf A}}_{{{\scriptstyle \rm supercharges}}}\,.
\eeq
The (anti)commutators in the three groups in \rf{19062026-man-25} are rewritten in the respective three lines as:
\beq
\label{19062026-man-05} && [M^{AB},M^{CE}] = \eta^{BC} M^{AE} + 3 \hbox{ terms},\qquad [M^{\isf\jsf},M^{\ksf\lsf}] = \epsilon^{\jsf\ksf}  M^{\isf\lsf} + 3 \hbox{ terms},
\nonumber\\
&& [M^{AB},M^{\isf C}] = \eta^{BC} M^{\isf A} - \eta^{AC} M^{\isf B}\,, \qquad  [M^{\isf\jsf},M^{\ksf A}] = \epsilon^{\jsf\ksf}  M^{\isf A} + \epsilon^{\isf\ksf}  M^{\jsf A}\,,
\nonumber\\
&& \hspace{2cm} \{ M^{A \isf },M^{\jsf B} \} =  \epsilon^{\isf\jsf} M^{AB} + \eta^{AB} M^{\isf\jsf}\,,
\eeq
$\epsilon^{12} = -\epsilon^{21}=1$. We note $M^{AB}=-M^{BA}$, $M^{\isf\jsf}=M^{\jsf\isf}$,  $M^{\isf A}=-M^{A\isf}$. Using the identification $X^1:=X^\eta$, $X^2: = X^\rho$, we represent \rf{19062026-man-30} as
\beq
&& [M^{AB}, X^C] = \eta^{BC} X^A - \eta^{AC} X^B\,, \qquad [M^{\isf\jsf} , X^\ksf ] = \epsilon^{\jsf\ksf}  X^\isf  + \epsilon^{\isf\ksf}  X^\jsf \,,
\nonumber\\
&& \{M^{A\isf},X^\jsf \} =  \epsilon^{\isf\jsf}  X^A\,, \hspace{3.2cm} [ M^{\isf A},X^B ] =  \eta^{AB}  X^\isf \,.
\eeq
The invariant scalar product of two vectors of the $osp(d,1|2)$ algebra given in \rf{19062026-man-35} is represented as:
\beq
\label{19062026-man-25scal} X\cdot Y =  X^A Y^A - \epsilon_{\isf\jsf} X^\isf Y^\jsf\,,\qquad  \qquad X^A Y^A: = \eta_{AB} X^A Y^B\,,
\eeq
$\epsilon_{12}=-\epsilon_{21}=1$.

\noinbf{Manifestly $osp(d,1|2)$ covariant basis}. Generators \rf{19062026-man-36} of the $osp(d,1|2)$ algebra are combined into $M^{\Ah\Bh}$, where  $\Ah,\Bh$ are vector indices of this algebra. These indices are decomposed as $\Ah=A,\isf$, $\Bh=B,\jsf$, where $A,B=0,1,\ldots,d$ are vector indices of the Lorentz algebra $so(d,1)$, while $\isf,\jsf=1,2$ are vector indices of the $sp(2)$ algebra. We therefore note that
\beq
\label{19062026-man-01} M^{\Ah\Bh} = (M^{AB}\,,  M^{\isf\jsf}\,, M^{\isf A})\,,\qquad M^{\Ah\Bh} = - (-)^{\varepsilon_\Ah\varepsilon_\Bh} M^{\Bh\Ah}\,,
\eeq
where we define the symbol $\varepsilon_\Ah$ by the following relations:
\beq
\label{19062026-man-10}  && \varepsilon_A = 0\,,  \qquad \varepsilon_\isf = 1 \,.
\eeq
The (anti)commutators \rf{19062026-man-05} can then be combined into the graded commutator given by:
\beq
\label{19062026-man-15}  && [M^{\Ah\Bh},M^{\Ch\Eh}\}= \eta^{\Bh\Ch} M^{\Ah\Eh} - (-)^{\varepsilon_\Ah\varepsilon_\Bh} \eta^{\Ah\Ch} M^{\Bh\Eh}
\nonumber\\
&& \hspace{2.4cm} -\, (-)^{\varepsilon_\Ch\varepsilon_\Eh} \eta^{\Bh\Eh} M^{\Ah\Ch} + (-)^{\varepsilon_\Ah\varepsilon_\Bh +\varepsilon_\Ch\varepsilon_\Eh} \eta^{\Ah\Eh} M^{\Bh\Ch}\,,
\eeq
where the metric tensor $\eta^{\Ah\Bh}$ and its cousin $\eta_{\Ah\Bh}$ are defined as:
{\small
\beq
\label{19062026-man-20} &&   \eta^{\Ah\Bh} = (\eta^{AB},\epsilon^{\isf\jsf})\,, \qquad \eta^{AB} = (-,+,\ldots,+), \qquad \epsilon^{12} = -\epsilon^{21}=1\,,
\nonumber\\
&&   \eta_{\Ah\Bh} = (\eta_{AB},\epsilon_{\isf\jsf})\,, \qquad \eta_{AB} = (-,+,\ldots,+), \qquad \epsilon_{12} = -\epsilon_{21}=1\,,
\nonumber\\
&& \eta^{\Ah\Bh} = (-)^{\varepsilon_{\Ah}} \eta^{\Bh\Ah}\,,\qquad \eta^{\Ah\Eh}\eta_{\Bh\Eh} = \delta_\Bh^\Ah\,, \qquad  \eta^{\Ah\Eh}\eta_{\Eh\Bh} = (-)^{\varepsilon_\Ah}\delta_\Bh^\Ah\,, \qquad
\eeq
}
\!and $\delta_\Ah^\Bh$ denotes the Kronecker delta. Using the notation $\varepsilon_X$ for the Grassmann parity of a quantity $X$, we recall the definition of the graded commutator:
\beq
&& [X,Y\} := XY - (-)^{\varepsilon_X\varepsilon_Y} YX\,,\qquad [X,Y\} = (-)^{\varepsilon_X\varepsilon_Y+1}[Y,X\}\,.
\eeq
Considering the vector $X^\Ah$ with $\varepsilon_{X^\Ah} = \varepsilon_{\Ah}$, we note the (anti)commutators:
\beq
&& [M^{\Ah\Bh}, X^\Ch\} = \eta^{\Bh\Ch} X^\Ah - (-)^{\varepsilon_\Ah\varepsilon_\Bh} \eta^{\Ah\Ch} X^\Bh\,.
\eeq
The invariant scalar product of two vectors of the $osp(d,1|2)$ algebra given in \rf{19062026-man-25scal} is represented as:
\beq
\label{19062026-man-20x} X\cdot Y = X^\Ah Y^\Ah\,, \qquad X^\Ah Y^\Ah : = \eta_{\Ah\Bh} X^\Bh Y^\Ah\,.
\eeq

\noinbf{Basis of $osp(d-1,1|2)$ subalgebra}. The generators $M^{\Ah\Bh}$ of the $osp(d,1|2)$ algebra are decomposed as:
\beq
M^{\Ah\Bh} = (M^{\ah\bh}, M^{z\ah})\,,
\eeq
where $M^{\ah\bh}$ are the generators of the $osp(d-1,1|2)$ subalgebra, while $M^{z\ah}$ are boost generators, $M^{\ah z}:= - M^{z\ah}$. (Anti)commutators of this subalgebra are obtained from \rf{19062026-man-15} by using the decomposition of the flat metric $\eta^{\Ah\Bh} = (\eta^{\ah\bh},\eta^{zz})$, where $\eta^{\ah\bh} = (\eta^{ab},\epsilon^{\isf\jsf})$, $\eta^{zz}=1$, while $\eta^{\Ah\Bh}$ is given in \rf{19062026-man-20}. The remaining (anti)commutators take the form:
\beq
&& [M^{z\ah},M^{\bh\ch}\} = \eta^{\ah\bh}M^{z\ch} - (-)^{\varepsilon_\bh\varepsilon_\ch} \eta^{\ah\ch} M^{z\bh} \,, \qquad  [M^{z\ah},M^{z\bh}] = - M^{\ah\bh}\,.
\eeq
Scalar product \rf{19062026-man-20x} is decomposed as $X^\Ah Y^\Ah = X^\ah Y^\ah + X^z Y^z$, where $X^\ah Y^\ah:= \eta_{\ah\bh} X^\bh Y^\ah$, $\eta_{\ah\bh} = (\eta_{ab}, \epsilon_{\isf\jsf})$.

\appendix{ \large Oscillator realization of $osp(d,1|2)$ algebra}\label{app-c}

\noinbf{\bf $so(d,1)\oplus sp_{\eta\rho}(2)$-basis}. We use the creation operators  $\alpha^A$, $\eta$, $\rho$ and the annihilation operators $\alphab^A$, $\rhob$, $\etab$, where their (anti)commutators are given in \rf{20062026-man-05}. The oscillator realization of the $osp(d,1|2)$ algebra is then given by the relations,
\beq
\label{21062026-man-01} && M^{AB} = \alpha^A\alphab^B - \alpha^B\alphab^A\,,\quad  M^{\eta\eta}  = 2\eta\etab\,,\quad M^{\rho\rho}  = -2\rho\rhob\,, \quad M^{\eta\rho}  =  N_\rho - N_\eta\,.
\nonumber\\
&& M^{\eta A} = \eta \alphab^A - \alpha^A \etab\,, \qquad M^{\rho A} = \rho \alphab^A + \alpha^A \rhob\,.
\eeq
The quantities $\alpha_\Tsm^A$, $\eta_\Tsm$, $\rho_\Tsm$ appearing in the operator $B^\Ah$ are defined as:
{\small
\beq
\label{21062026-man-10} && \alpha_\Tsm^A = \alpha^A - (\alpha^B\alpha^B + 2 \rho\eta)\frac{1}{2N_\alphabfwh  + d-1} \alphab^A\,,\qquad N_\alphabfwh: = N_\alpha + N_\rho + N_\eta\,,
\nonumber\\
&& \eta_\Tsm = \eta  - (\alpha^B\alpha^B + 2 \rho \eta) \frac{1}{2N_\alphabfwh + d-1} \etab\,, \hspace{1cm} \rho_\Tsm = \rho  + (\alpha^B\alpha^B + 2 \rho \eta) \frac{1}{2N_\alphabfwh + d-1} \rhob\,.\qquad
\eeq
}

\noinbf{\bf $so(d,1)\oplus sp(2)$-basis}. We use the creation operators  $\alpha^A$, $\alpha^\isf$ and the annihilation operators $\alphab^A$, $\alphab^i$. The oscillator realization of the $osp(d,1|2)$ algebra is then given by the relations,
\beq
\label{21062026-man-20} && M^{AB} = \alpha^A\alphab^B - \alpha^B \alphab^A\,, \qquad M^{\isf\jsf}  = \alpha^\isf \alphab^\jsf  + \alpha^\jsf  \alphab^\isf \,,\qquad M^{\isf A} = \alpha^\isf \alphab^A - \alpha^A \alphab^\isf \,,\qquad
\nonumber\\
&& [\alphab^A,\alpha^B] = \eta^{AB}\,,\hspace{2cm} \{\alphab^\isf ,\alpha^\jsf \} = \epsilon^{\isf\jsf} \,.
\eeq
In the basis under consideration, the quantities in \rf{21062026-man-10} are represented as
{\small
\beq
\label{21062026-man-25} && \alpha_\Tsm^A := \alpha^A - (\alpha^B\alpha^B + \epsilon_{\jsf\ksf}\alpha^\ksf\alpha^\jsf) \frac{1}{2N_\alphabfwh + d-1}\alphab^A\,,
\nonumber\\
&& \alpha_\Tsm^\isf := \alpha^\isf - (\alpha^B\alpha^B + \epsilon_{\jsf\ksf} \alpha^\ksf \alpha^\jsf) \frac{1}{2N_\alphabfwh + d-1} \alphab^\isf\,,\qquad N_\alphabfwh = \alpha^A\alphab^A + \varepsilon_{\isf\jsf}\alpha^\jsf \alphab^\isf\,.\qquad
\eeq
}
\!The Grassmann-odd oscillators $\alpha^\isf$, $\alphab^\isf$ and $\eta$, $\rho$, $\etab$, $\rhob$ are identified as:
\beq
&& \hspace{1cm} \alpha^1 = \eta, \qquad \alpha^2 = \rho,\qquad \alphab^1 = \etab,\qquad \alphab^2 = -\rhob\,.
\eeq

\noinbf{Manifestly $osp(d,1|2)$ covariant basis}. Now the oscillator realization takes the form:
\beq
\label{21062026-man-50} && M^{\Ah\Bh} = \alpha^\Ah \alphab^\Bh - (-)^{\varepsilon_\Ah\varepsilon_\Bh}\alpha^\Bh \alphab^\Ah\,, \qquad \qquad [\alphab^\Ah,\alpha^\Bh\} = \eta^{\Ah\Bh}\,.
\eeq
In terms of the Grassmann-even and Grassmann-odd oscillators, the oscillators $\alpha^\Ah$, $\alphab^\Ah$ are decomposed as:
\beq
&& \alpha^\Ah = \alpha^A\,,\ \ \eta\,, \ \ \rho\,,\qquad \alphab^\Ah = \alphab^A\,,\ \etab\,, -\rhob\,.
\eeq
In the basis under consideration, the quantities in \rf{21062026-man-25} are combined into $\alpha_\Tsm^\Ah$, which is given by:
\beq
\label{21062026-man-55} && \alpha_\Tsm^\Ah := \alpha^\Ah - \alpha^\Bh \alpha^\Bh \frac{1}{2N_\alphabfwh + d-1}\alphab^\Ah\,,\qquad N_\alphabfwh := \alpha^\Ah \alphab^\Ah\,.
\eeq

\noinbf{Useful relations for $\alpha_\Tsm^\Ah$}. Let $\Phi$ be an $osp(d,1|2)$-traceless field, $\alphab^\Ah\alphab^\Ah\Phi=0$. Using the shorthand $A\approx 0$ instead of $A\Phi=0$, we note the following strong and weak relations:
{\small
\beq
&&  \hspace{-0.8cm} \alpha_\Tsm^\Ah \alpha_\Tsm^\Bh \approx (-)^{\varepsilon_\Ah\varepsilon_\Bh} \alpha_\Tsm^\Bh \alpha_\Tsm^\Ah \,,\hspace{0.5cm} \alpha_\Tsm^\Ah \alpha_\Tsm^\Ah \approx 0\,,\hspace{1cm} \alphab^\Bh \alphab^\Bh \alpha_\Tsm^{\Ah_1}\ldots \alpha_\Tsm^{\Ah_n}\approx 0\,,
\nonumber\\
&& \hspace{-0.8cm} 2\alpha_\Tsm^{[\Ah} \alphab^{\Bh\}} = M^{\Ah\Bh}\,,\hspace{1.5cm}  2\alphab^{[\Ah} \alpha_\Tsm^{\Bh\}} = - \frac{2N_\alphabfwh +
d-1}{2N_\alphabfwh+d-3} M^{\Ah\Bh}\,,
\nonumber\\
&& \hspace{-0.8cm} \alphab^\Ah \alpha^\Ah = N_\alphabfwh + d-1\,, \hspace{1cm} \alphab^\Ah \alpha_\Tsm^\Ah \approx \frac{(N_\alphabfwh + d-3)(2N_\alphabfwh + d-1)}{2N_\alphabfwh + d-3}\,,\quad \alpha_\Tsm^\Ah\alphab^\Ah \approx N_\alphabfwh\,, \qquad
\nonumber\\
&& \hspace{-0.8cm}  M^{\Ah\Bh} \alphab^\Bh \approx - N_\alphabfwh \alphab^\Ah\,,\hspace{1.2cm} M^{\Ah\Bh} \alpha_\Tsm^\Bh \approx \alpha_\Tsm^\Ah (N_\alphabfwh + d-2)\,,
\eeq
}
where $ X^{[\Ah} Y^{\Bh\}} := \half X^{\Ah} Y^{\Bh} - \half(-)^{\varepsilon_\Ah\varepsilon_\Bh} X^\Bh Y^\Ah$.

\appendix{ \large Matching metric-like Lagrangian and BRST-BV Lagrangian}\label{app-d}

The realization of a gauge invariant Lagrangian for massless, massive, partially-massless, and con\-tinuous-spin fields in terms of the standard de Donder derivative was obtained in Refs.\cite{Metsaev:2008ks,Metsaev:2009hp,Metsaev:2016lhs}. Here, to treat all fields under consideration on an equal footing, we slightly update the notation used in those papers. This is to say that, in $AdS_{d+1}$ space,  the gauge-invariant Lagrangian for above-mentioned fields can be presented on an equal footing as follows:
{\small
\beq
\label{18062026-man-01} &&\hspace{-1cm} e^{-1}\LL  =   \Phi_\cvrm^\dagger  E \Phi_\cvrm \,,
\nonumber\\
&& E := (1-\frac{1}{4}\alphabf^2 \bar\alphabf^2) \big(\Box_{_{AdS}} - \CC_2 + N_\alphabf(N_\alphabf + d-1) -
\alphabf^2\bar\alphabf^2\big) - \Cbf_\st\bar\Cbf_\st\,,
\nonumber\\
&& \bar\Cbf_\st :=  \bar\alphabf \Dbf - \half \alphabf \Dbf \bar\alphabf^2 +
\Pibf\bar\bbf + \half \bbf \bar\alphabf^2\,,\hspace{1cm} \Box_{_{AdS}}=D^A D^A + \omega^{AAB}D^B\,,
\nonumber\\
&& \Cbf_\st : =  \alphabf \Dbf  - \half \alphabf^2 \bar\alphabf \Dbf - \bbf
\Pibf - \half \alphabf^2\bar\bbf\,,\qquad \Pibf := 1 - \alphabf^2\frac{1}{2(2N_\alphabf +
d+1)}\bar\alphabf^2\,,\quad
\nonumber\\
&& \alphabf \Dbf:= \alpha^A D^A\,, \quad \bar\alphabf\, \Dbf:= \alphab^A D^A\quad \alphabf^2:= \alpha^A \alpha^A\,, \quad \bar\alphabf^2:= \alphab^A \alphab^A\,,
\nonumber\\
&& D^A = e^{\mun A} D_\mun\,, \quad D_\mun = \partial_\mun + \half \omega_\mun^{AB}M^{AB}\,,\quad\omega^{ABC}:=e^{\mun A}\omega_\mun^{BC}\,,\quad  e = \det e_\mun^A\,,\qquad
\eeq
}
\!where $\Phi_\cvrm$ is the double traceless field, $(\bar\alphabf^2)^2\Phi_\cvrm=0$ and $e_\mun^A$ is the vielbein.

\noinbf{\ibf)} For massless, massive, partially-massless fields, and continuous-spin field, the field $\Phi_\cvrm$ is represented in terms of the component fields in the same way as the field $\phi_0$ in \rf{03062026-man-10}, \rf{07062026-man-05}, \rf{08062026-man-05}.
For a massless field $\bbf=0$, $\bar\bbf=0$, while for massive, partially-massless and continuous-spin fields, these operators are given in \rf{07062026-man-15}, \rf{07062026-man-16}
and \rf{08062026-man-25} respectively.%
\footnote{For a massless field in $AdS_4$, the Lagrangian \rf{18062026-man-01} coincides with the Fronsdal Lagrangian. For $d>3$, the Fronsdal operator was obtained in Ref.\cite{Metsaev:1999ui}. In terms of the Fronsdal operator denoted as $F$, Lagrangian \rf{18062026-man-01} takes the form $e\Phi_\cvrm^\dagger (1-\frac{1}{4}\alphabf^2\bar\alphabf^2)F\Phi_\cvrm$.}

\noinbf{\iibf)} For a massless field, the operator $\bar\Cbf_\st$ is well known, while for massive, partially-massless and continuous-spin fields, the contribution of the operators $\bbf$, $\bar\bbf$ to $\bar\Cbf_\st$ was found in Refs.\cite{Metsaev:2009hp,Metsaev:2016lhs}. The differential operators $ \Cbf_\st$ and $\bar\Cbf_\st$ are related via $(e\Cbf_\st)^\dagger = - e \bar\Cbf_\st$. Hence, up to a total derivative, we obtain the relation $e\Phi_\cvrm^\dagger \Cbf_\st \bar\Cbf_\st \Phi_\cvrm = - e L_\st \Lb_\st$, where $\Lb_\st = \bar\Cbf_\st\Phi_\cvrm$, $L_\st:=\Lb_\st^\dagger$. Since the operator $\bar\Cbf_\st$  is written in terms of the standard de Donder derivative, we say that \rf{18062026-man-01} provides a realization of the Lagrangian in terms of the standard  de Donder derivative.

To rewrite the Lagrangian \rf{18062026-man-01} in terms of the modified de Donder derivative, we use the Poincar\'e parametrization of AdS space. Then, as shown in Refs.\cite{Metsaev:2008ks,Metsaev:2009hp}, the gauge-invariant Lagrangian \rf{18062026-man-01} can be represented as:
\beq
\label{18062026-man-50} && e^{-1} \LL = \Phi_\cvrm^\dagger\big( \Box_{_{0\,AdS}} - A_\cvrm\big) \Phi_\cvrm  - \frac{1}{4}\Phi_\cvrm^\dagger\alphabf^2 \big( \Box_{_{0\,AdS}}  - A_\cvrm^\alphabf\big) \bar\alphabf^2\Phi_\cvrm - \Phi_\cvrm^\dagger \Cbf_\md \bar\Cbf_\md \Phi_\cvrm\,,
\nonumber\\
&& \hspace{1cm} \Box_{_{0\,AdS}} =  z^2(\Box+\partial_z^2) +(1-d)z\partial_z \,,
\nonumber\\
&& \hspace{1cm} A_\cvrm : = \CC_2 - N_\alphabf(N_\alphabf + d-2) -  N_{\alpha^z} (2N_\alphabf + d - 5)
+ 2 \alpha^z \bar\bbf + 2 \bbf  \bar\alpha^z\,,
\nonumber\\
&& \hspace{1cm} A_\cvrm^\alphabf := \CC_2 - (N_\alphabf+2)(N_\alphabf + d) -  N_{\alpha^z} (2N_\alphabf + d - 1) + 2 \alpha^z \bar\bbf + 2 \bbf  \bar\alpha^z\,, \qquad
\nonumber\\
&& \hspace{1cm}  \Cbf_\md : =  \Cbf_\st - 2 \Cbf_\perp^z \,,\hspace{2.4cm} \bar\Cbf_\md := \bar\Cbf_\st + 2\bar{\Cbf}_\perp^z \,,
\nonumber\\
&& \hspace{1cm}  \Cbf_\perp^z \equiv \alpha^z  - \half \alphabf^2
\bar\alpha^z \,, \hspace{2.5cm}
\bar\Cbf_\perp^z \equiv \bar \alpha^z -\half \alpha^z \bar\alphabf^2 \,.
\eeq
In view of the $\bar{\Cbf}_\perp^z$-term in $\bar\Cbf_\md$ we say that $\bar\Cbf_\md$ is written in terms of the modified de Donder derivative, while \rf{18062026-man-50} provides a realization of the Lagrangian in terms of the modified de Donder derivative.
Using the relation $(e\Cbf_\md)^\dagger = - e \bar\Cbf_\md$, we find that, up to total derivative,  $e\Phi_\cvrm^\dagger \Cbf_\md \bar\Cbf_\md \Phi_\cvrm = - e L_\md \Lb_\md$, where $\Lb_\md = \bar\Cbf_\md\Phi_\cvrm$, $L_\md:=\Lb_\md^\dagger$. In the Poincar\'e parametrization of AdS, the derivative $D^A$ \rf{18062026-man-01} takes the form $D^A = z \partial^A + M^{zA}$, $\partial^A = \eta^{AB}\partial/\partial x^B$, $M^{zA} = \alpha^z\alphab^A - \alpha^A\alphab^z$, $x^A:=\delta_\mun^A x^\mun$.

\noinbf{BRST-BV Lagrangian}. To match our BRST-BV Lagrangian \rf{02062026-man-30} with the metric-like one we must eliminate fields of the BRST-BV Lagrangian that have non-zero ghost number. From \rf{20062026-man-55}, we see that the fields $\phi_\eta$, $\phi_\rho$, $\phi_{*0}$, $\phi_{*\eta}$, and $\phi_{*\eta\rho}$ have non-zero ghost number; among these the fields $\phi_\eta$, $\phi_{*0}$, $\phi_{*\eta\rho}$ can be gauged away by using the gauge transformations:
\beq
\label{18062026-man-65gaug} && \phi_\eta=0,\qquad  \phi_{*0}=0, \qquad \phi_{*\eta\rho}=0\,.
\eeq
The remaining fields with non-zero ghost numbers $\phi_\rho$, $\phi_{*\eta}$ cannot be gauged away. However, upon applying gauge conditions \rf{18062026-man-65gaug}, their contribution to the BRST-BV Lagrangian cancels out.
We are then left with the BRST-BV Lagrangian that depends on the fields having zero ghost number, $\phi_0$, $\phi_{\eta\rho}$,  $\phi_{*\rho}$. In view of the tracelessness constraint, $\phi_{\eta\rho}$ is traceless and can be expressed in terms of $\phi_0$, which turns out to be double-traceless. Furthermore, using equations of motion for $\phi_{*\rho}$, we can express  $\phi_{*\rho}$ in terms of $\phi_0$. As a result, we obtain the relations
\beq
\label{18062026-man-67} && (\bar\alphabf^2)^2\phi_0=0\,, \quad \phi_{\eta\rho}= \half \bar\alphabf^2\phi_0\,, \quad \phi_{*\rho} = \Lb\,, \qquad z \Lb := (\bar\Cbf_\md + \frac{d-1}{2} \bar\Cbf_\perp^z)\phi_0\,,
\eeq
where $\bar\Cbf_\md$, $\Cb_\perp^z$ are defined in \rf{18062026-man-01}, \rf{18062026-man-50}.
Using \rf{18062026-man-67} in the BRST-BV Lagrangian \rf{02062026-man-30}, we find
\beq
\label{18062026-man-70} && \LL = \phi_0^\dagger\big( \Box +\partial_z^2 - \frac{1}{z^2} A\big) \phi_0  - \frac{1}{4} \phi_0^\dagger\alphabf^2 \big( \Box +\partial_z^2  - \frac{1}{z^2} A^\alphabf\big) \bar\alphabf^2 \phi_0 + L\Lb\,,
\nonumber\\
&& \hspace{1cm} A: = A_\cvrm + \frac{d^2-1}{4}\,, \qquad A^\alphabf  = A_\cvrm^\alphabf + \frac{d^2-1}{4}\,,\qquad L = \Lb^\dagger\,,
\eeq
where $A_\cvrm$, $A_\cvrm^\alphabf$ are defined in \rf{18062026-man-50}.

Finally, we note that $\Phi_\cvrm$ is a covariant field, while $\phi_0$ in \rf{03062026-man-10}, \rf{07062026-man-05}, \rf{08062026-man-05} is a canonically normalized field. These fields are related as
\be \label{18062026-man-59}
\Phi_\cvrm = z^{ \frac{d-1}{2} } \phi_0\,.
\ee
Plugging \rf{18062026-man-59} into Lagrangian $\LL$ \rf{18062026-man-50}, we find that the Lagrangian $\LL$ \rf{18062026-man-50} coincides exactly with the one derived from BRST-BV formulation \rf{18062026-man-70}.

For massless, massive, and partially-massless fields, operators $A$, $A^\alphabf$ \rf{18062026-man-70} are simplified
{\small
\beq
\label{18062026-man-80}&& \hspace{-2.7cm} A = (s+ \frac{d-3}{2})(s+ \frac{d-5}{2}) - N_{\alpha^z}(2s + d-5)\,, \hspace{1cm} A^\alphabf = A\,, \hspace{0.4cm} \hbox{for massless};
\nonumber\\
&&\hspace{-2.7cm} A = m^2 + (s+ \frac{d-3}{2})(s+ \frac{d-5}{2}) - N_{\alpha^z}(2s + d-5- 2 N_\zeta)
\nonumber\\
&& \hspace{-2.2cm} +\,\, N_\zeta (2s+d-2 - N_\zeta) + 2 \alpha^z \bar\bbf + 2 \bbf  \bar\alpha^z\,, \hspace{0.4cm} A^\alphabf = A\,,  \hspace{0.3cm} \hbox{for massive/partial-massless};
\eeq
}
where $\bbf$, $\bar\bbf$  are given in \rf{07062026-man-15}, \rf{07062026-man-16}, while $m^2$ for the partially-massless field is given below in \rf{18062026-man-90}. In deriving \rf{18062026-man-80}, we use \rf{18062026-man-50}, \rf{18062026-man-70}, and homogeneity conditions \rf{03062026-man-15}, \rf{07062026-man-07}.

\noinbf{Casimir operator, energy and mass parameters}. Here, for the reader convenience we collect our notation and conventions. The eigenvalues of the 2nd- and 4th-order Casimir operators of the $so(d,2)$ algebra are given by
\beq
\label{18062026-man-65} \CC_2 = E_0(E_0-d) + s(s+d-2)\,, \qquad \CC_4 = (E_0-1)(E_0-d+1)s(s+d-2)\,,
\eeq
where $E_0$ and $s$ are the respective energy and spin parameters. One has the following relations for the irreps associated with massless, massive, and partially-massless fields:
{\small
\beq
\label{18062026-man-90} && \hspace{-1cm} E_0 = s + d-2\,,\hspace{3.5cm} \hbox{for massless}, s\geq 1\,;
\nonumber\\
&& \hspace{-1cm} E_0= \frac{d}{2} + \sqrt{m^2 + \big(s + \frac{d-4}{2}\big)^2} \,, \qquad \hbox{for massive, $s\geq 1$, and partially-massless, $s\geq 2$};
\nonumber\\
&& \hspace{-1cm} E_0 = s + d - 2 -t \,,\qquad m^2 = - t (2s+ d-4-t)\,, \hspace{0.3cm} \hbox{for partially-massless}, s\geq 2\,.\qquad
\eeq
}
\!For continuous-spin field, we prefer to use the labels $p$ and $q$ and hence the following  values of $\CC_2$ and $\CC_4$,
{\small
\beq
 \label{18062026-man-75} && p: = E_0 - \frac{d}{2}\,, \qquad q := s + \frac{d-2}{2}\,,
\nonumber\\
&& \CC_2 = p^2 + q^2 - \frac{d^2+(d-2)^2}{4}\,, \qquad \CC_4 = \big(p^2 - \frac{(d-2)^2}{4}\big)\big(q^2 - \frac{(d-2)^2}{4}\big)\,.\qquad
\eeq
}

\appendix{ \large Spin operators of unconstrained formulation }\label{app-e}

\noinbf{Spin operators and operator $B^A$ for unconstrained formulation with $K\ne 1$}.
\beq
\label{23062026-man-01} && K  :=  k_0 (1-N_\chi) + N_\chi k_1\,,
\nonumber\\
&& k_0 = \sum_{n=0}^\infty \frac{(\alpha^\ah\alpha^\ah)^n(\alphab^\bh\alphab^\bh)^n}{(2n)!}\,, \qquad k_1 = \sum_{n=0}^\infty \frac{(\alpha^\ah\alpha^\ah)^n(\alphab^\bh\alphab^\bh)^n}{(2n+1)!}\,,
\\[15pt]
\label{23062026-man-05} &&  \hspace{-1cm} M^{\ah\bh}=\alpha^\ah \alphab^\bh - (-)^{\varepsilon_\ah\varepsilon_\bh}\alpha^\bh \alphab^\ah \,,
\nonumber\\
&&  \hspace{-1cm} M^{z\ah}=\chi (\alphab^\ah  +\alpha^\ah \alphab^\bh\alphab^\bh)- \alpha^\ah \chib\,, \qquad M^{\ah z} := - M^{z\ah}\,,
\\[10pt]
&& \hspace{-1cm} B^\Ah = \bbf\alphab_\chi^\Ah + \alpha_{\Tsm\chi}^\Ah \bar\bbf\,, \qquad B^\Ah = (B^\ah,B^z)\,,
\nonumber\\
\label{23062026-man-10} && \alpha_{\Tsm\chi}^\ah  = \alpha^\ah  - \alpha^\bh\alpha^\bh \frac{1}{2N_\alphawh + 2N_\chi + d-1}\alphab^\ah \,,
\nonumber\\
&& \alpha_{\Tsm\chi}^z = \chi \Big(1 + \alpha^\ah\alpha^\ah \frac{1}{2N_\alphawh+d+1} \alphab^\bh\alphab^\bh \Big) -  \alpha^\ah\alpha^\ah \frac{1}{2N_\alphawh+d-1}\chib\,,
\nonumber\\
&& \alphab_\chi^\ah  = \alphab^\ah \,, \qquad  \alphab_\chi^z = \chib - \chi\alphab^\ah\alphab^\ah\,.
\eeq

\noinbf{Spin operators and operator $B^A$ for unconstrained formulation with $K = 1$}.
\beq
\label{23062026-man-30} && \hspace{-1cm} M^{\ah\bh}=\alpha^\ah \alphab^\bh -(-)^{\varepsilon_\ah\varepsilon_\bh} \alpha^\bh \alphab^\ah \,,
\nonumber\\
&& \hspace{-1cm} M^{z\ah} = \chi \alphab_{10}^\ah  - \alpha_{01}^\ah \chib\,, \quad\alpha_{01}^\ah  = k_0^{1/2}\alpha^\ah  k_1^{-1/2}\,,\quad  \alphab_{10}^\ah  = k_1^{-1/2}\alphab^\ah k_0^{1/2}\,,\quad M^{\ah z} := - M^{z\ah}\,, \qquad
\\[10pt]
\label{23062026-man-35} && \hspace{-1cm} B^\Ah = \bbf \alphab_\Ksm^\Ah + \alpha_\Ksm^\Ah \bar\bbf\,,\qquad B^\Ah = (B^\ah,B^z)\,,
\nonumber\\
&& \alpha_\Ksm^\ah  = (1-N_\chi) \alpha_{00}^\ah    + N_\chi \alpha_{11}^\ah  \,, \hspace{1cm} \alphab_\Ksm^\ah  = (1-N_\chi) \alphab_{00}^\ah   + N_\chi \alphab_{11}^\ah \,,
\nonumber\\
&& \alpha_\Ksm^z =   \chi k_{10}^{\vphantom{5pt}} + \alpha_{01}^{\vphantom{5pt}} \chib   \,,\hspace{2.5cm} \alphab_\Ksm^z = k_{01}^{\vphantom{5pt}} \chib  +  \chi \alphab_{10}^{\vphantom{5pt}}\,,
\nonumber\\[15pt]
&& \alpha_{00}^\ah := k_0^{-1/2} \alpha^\ah  k_0^{1/2}\,, \hspace{2.5cm} \alphab_{00}^\ah  :=k_0^{1/2} \alphab^\ah  k_0^{-1/2}\,,
\nonumber\\
&& \alpha_{11}^\ah := k_1^{-1/2}\alpha^\ah  k_1^{1/2}\,, \hspace{2.5cm} \alphab_{11}^\ah := k_1^{1/2}\alphab^\ah  k_1^{-1/2}\,,
\nonumber\\
&& \alpha_{01}^{\vphantom{5pt}} := - k_0^{-1/2}\alpha^2 k_1^{1/2}\,,\hspace{2.2cm} \alphab_{10}^{\vphantom{5pt}}:= - k_1^{1/2}\alphab^2 k_0^{-1/2}\,,
\nonumber\\
&& k_{10}^{\vphantom{5pt}} := k_1^{-1/2} k_0^{1/2}\,, \hspace{3cm} k_{01}^{\vphantom{5pt}} := k_0^{1/2} k_1^{-1/2}\,,
\eeq
where operators $k_0$, $k_1$ are given in \rf{23062026-man-01}.

\noinbf{Comment on hermiticity properties of spin operators}. Using \rf{23062026-man-01}-\rf{23062026-man-10}, we find that hermiticity conditions \rf{19062026-man-37} amount to the relations
\beq
\label{23062026-man-37} &&  k_0 \alpha^\ah  = (\alpha^\ah + \alpha^\bh\alpha^\bh\alphab^\ah)k_1\,,
\\
\label{23062026-man-38} && k_0 \Big(\alpha^\ah - \alpha^\bh\alpha^\bh \frac{1}{2N_\alphawh+d-1}\alphab^\ah\Big)  = \alpha^\ah k_0\,,
\nonumber\\
&& k_1 \Big(\alpha^\ah - \alpha^\bh\alpha^\bh \frac{1}{2N_\alphawh+d+1}\alphab^\ah\Big)  = \alpha^\ah k_1\,,
\nonumber\\
&& k_0 \alpha^\bh\alpha^\bh \frac{1}{2N_\alphawh+d-1} = \alpha^\bh\alpha^\bh k_1 \,,
\nonumber\\
&&  k_1\Big( 1 + \alpha^\bh\alpha^\bh \frac{1}{2N_\alphawh+d+1}\alphab^\ch\alphab^\ch\Big) = k_0 \,,
\eeq
where relation \rf{23062026-man-37} is obtained from hermiticity condition for $M^{z\ah}$, while relations \rf{23062026-man-38} are obtained from the one for $B^\Ah$. We directly checked that relations \rf{23062026-man-37}, \rf{23062026-man-38} hold true.

\noinbf{Comments on derivation of relations \rf{23062026-man-01}-\rf{23062026-man-10}}.
Following the framework in App.~B of Ref.\cite{Metsaev:1999ui}, we now briefly comment on the derivation of relations \rf{23062026-man-01}-\rf{23062026-man-10}. Treating the $osp(d,1|2)$-tracelessness constraint $\alphab^\Ah\alphab^\Ah\Phi = 0$ as the 2nd-order differential equation with respect to the oscillator $\alpha^z$, we can represent the field $\Phi$ as
\beq
\label{23062026-man-50} && \Phi(\alpha^\ah,\alpha^z) = \cos(\omega \alpha^z) \Phi_0(\alpha^\ah) +  \frac{\sin(\omega \alpha^z)}{\omega} \Phi_1(\alpha^\ah)\,, \qquad \omega^2:= \alphab^\ah\alphab^\ah\,,
\eeq
where $\Phi_0$ and $\Phi_1$ are constraint-free fields depending on the oscillator $\alpha^\ah$ which is a vector under the $osp(d-1,1|2)$ subalgebra. In \rf{23062026-man-50} and below, we hide the dependence on the oscillators $\zeta$ and $\upsilon$ which are required when studying massive, partially-massless and continuous-spin fields. Since these oscillators do not appear in the tracelessness constraint, they are immaterial to our discussion. Using a new Grassmann-even oscillator $\chi$, $\chi^2=0$, we combine the fields $\Phi_0$ and $\Phi_1$ into a new field $\Phi_\chi$ given by
\beq
\label{23062026-man-55} \Phi_\chi(\alpha^\ah,\chi) = \Phi_0(\alpha^\ah) + \chi \Phi_1(\alpha^\ah)\,.
\eeq
We then note:

\noinbf{i)} The action of spin operators $M^{\ah\bh}$ and $M^{z\ah}$ \rf{21062026-man-50} on  field $\Phi$ \rf{23062026-man-50} amounts to the action of spin operators $M^{\ah\bh}$  and $M^{z\ah}$  \rf{23062026-man-05} on field $\Phi_\chi$ \rf{23062026-man-55}.

\noinbf{ii)} The action of operators $\alpha_\Tsm^\ah$, $\alpha_\Tsm^z$, $\alphab^\ah$, and $\alphab^z$ \rf{21062026-man-55} on field $\Phi$ \rf{23062026-man-50} amounts to the action of operators $\alpha_{\Tsm\chi}^\ah$, $\alpha_{\Tsm\chi}^z$, $\alphab_\chi^\ah$, and $\alphab_\chi^z$  \rf{23062026-man-10} on field $\Phi_\chi$ \rf{23062026-man-55}.

\noinbf{iii)} Using \rf{23062026-man-50}, we find that the scalar product of $\Phi$ can be presented in terms of $\Phi_\chi$ as
\beq
\label{23062026-man-60} \Phi^\dagger \Phi = \Phi_\chi^\dagger K \Phi_\chi\,,
\eeq
where $K$ is given in \rf{23062026-man-01}. The appearance of $K$ in \rf{23062026-man-60} explains the use of $K$ in the unconstrained formulation with $K\ne 1$.

As a side remark, the action of operator $N_\alphabfwh$ \rf{20062026-man-10} on field \rf{23062026-man-50} amounts to the action of the operator $N_\alphawh + N_\chi$ on field $\Phi_\chi$ \rf{23062026-man-55}. Therefore, when passing from the constrained formulation to the unconstrained formulations, homogeneity conditions \rf{03062026-man-15}, \rf{07062026-man-07}, \rf{08062026-man-07}
take the form given in \rf{03062026-man-15ax}, \rf{03062026-man-15a}, and \rf{08062026-man-60}, respectively.

In the main body of the paper, field  $\Phi_\chi$ \rf{23062026-man-55} is simply denoted as $\Phi$.

\noinbf{Comments on relations \rf{23062026-man-30}-\rf{23062026-man-35}}. The unconstrained formulation with $K=1$ is simply obtained by introducing a new field $\Phi_\Ksm$ via the following similarity transformation:
\be \label{23062026-man-70}
\Phi_\chi = K^{-1/2} \Phi_\Ksm \,, \qquad \Phi^\dagger K \Phi_\chi = \Phi_\Ksm^\dagger \Phi_\Ksm\,.
\ee
We then note:

\noinbf{i)} The action of spin operators $M^{\ah\bh}$ and $M^{z\ah}$ \rf{23062026-man-05} on  field $\Phi_\chi$ \rf{23062026-man-55} amounts to the action of spin operators $M^{\ah\bh}$  and $M^{z\ah}$ \rf{23062026-man-30} on field $\Phi_\Ksm$ \rf{23062026-man-70}.

\noinbf{ii)} The action of operators $\alpha_{\Tsm\chi}^\ah$, $\alpha_{\Tsm\chi}^z$, $\alphab_\chi^\ah$, and $\alphab_\chi^z$  \rf{23062026-man-10} on field $\Phi_\chi$ \rf{23062026-man-55} amounts to the action of operators $\alpha_\Ksm^\ah$, $\alpha_\Ksm^z$, $\alphab_\Ksm^\ah$, and $\alphab_\Ksm^z$  \rf{23062026-man-35}  on field $\Phi_\Ksm$ \rf{23062026-man-70}.

In the main body of the paper, field  $\Phi_\Ksm$ \rf{23062026-man-70} is simply denoted as $\Phi$.

\section{ Method of $M\hbox{-}T$ pair for solving equations \rf{02062026-man-50}}

In App.C of Ref.\cite{Metsaev:2004ee}, we proposed a method for solving the light-cone gauge counterparts of the equations \rf{02062026-man-50}. Following that framework, we now generalize  this method to the equations \rf{02062026-man-50} themselves. Our framework is applicable for studying fields in AdS with arbitrary spin and any type of symmetry. We begin with the following ansatz for the operator $B^\Ah$:
{\small
\be \label{05072026-man-01}
B^\Ah : = p\, T^\Ah + \half (M^{\Ah \Bh},T^\Bh\}\,, \qquad (M^{\Ah \Bh},T^\Bh\} : = M^{\Ah\Bh}T^\Bh - T^\Bh M^{\Bh\Ah}\,,\quad p:=E_0-\frac{d}{2}\,,\qquad
\ee
}
where we introduce the vector $T^\Ah$ of the $osp(d,1|2)$ algebra, which must satisfy the equations:
\beq
\label{05072026-man-05} [T^\Ah,T^\Bh\} = M^{\Ah\Bh}\,.
\eeq
Using \rf{05072026-man-05}, we find
\beq
\label{05072026-man-10} && [B^\Ah, B^\Bh\}= \Big(T^\Ch T^\Ch + p^2 + \frac{(d-2)(d-4)}{4}\Big)M^{\Ah\Bh} - (M^3)^{[\Ah|\Bh\}}\,.
\eeq
Comparing \rf{05072026-man-10} with the basic equations \rf{02062026-man-50}, we obtain the equations
\beq
\label{05072026-man-15} T M^{\Ah\Bh}=0\,, \qquad T:= T^\Ah T^\Ah + \half M^{\Ah\Bh}M^{\Bh\Ah} - \CC_2 + p^2 - \frac{d^2}{4}\,.
\eeq
Note that, if $M^{\Ah\Bh}\Phi\ne 0$ for all $\Phi$, then equations \rf{05072026-man-15} reduce to the single equation $T=0$. Usually the operator $M^{\Ah \Bh}$ is determined by  group-theoretic methods. The problem is therefore reduced to finding $T^\Ah$. We now summarize our method in the following three steps:

\noinbf{i)} Finding a general expression for $T^\Ah$ that satisfies equations \rf{05072026-man-05}.

\noinbf{ii)} Solving constraints imposed by equations \rf{05072026-man-15} on the general solution for $T^\Ah$.

\noinbf{iii)} Substituting the solution for $T^\Ah$ into relation \rf{05072026-man-01} and determining restrictions arising from the hermiticity of $B^\Ah$: $B^{A\dagger}=B^A$,  $B^{\eta \dagger}=B^\eta$,  $B^{\rho\dagger}=-B^\rho$.

These steps are implemented in the same way as in the light-cone gauge analysis in App.C of Ref.\cite{Metsaev:2004ee}. Therefore, to avoid repetition, we omit the details of the derivation of $B^\Ah$.

The solution for $B^\Ah$ presented in this paper was found by solving basic equations \rf{02062026-man-50} directly, as well as by using the method of the $M\hbox{-}T$ pair presented above. The results obtained from both methods of solving equations \rf{02062026-man-50} coincide. Finally, we anticipate that the method of the $M\hbox{-}T$ pair can also be used to find $B^\Ah$ corresponding to mixed-symmetry fields. All the equations given above are valid for these fields.

\end{document}